\hoffset -1 true cm
\overfullrule=0 pt
\raggedbottom
\newcount\equationno      \equationno=0
\newtoks\chapterno \xdef\chapterno{}
\newdimen\tabledimen  \tabledimen=\hsize
%
%
%

\def\eqname#1{\global \advance \equationno by 1 \relax
\xdef#1{{\noexpand{\rm}(\chapterno\number\equationno)}}#1}
%
%
%
%
\def\tablet#1#2{
\vbox{\tabskip=1em plus 4em minus 0.9em
\halign to #1{#2}} }
%
\def\tabmidrule{\noalign{\smallskip\hrule\smallskip}}             
\input epsf
%
%
%
%

\catcode `\@=11 

\def\@version{1.6}
\def\@verdate{18th September 1995}

%
%


\newif\ifprod@font

\ifx\@typeface\undefined
  \def\@typeface{Comp. Modern}\prod@fontfalse
\else
  \prod@fonttrue 
\fi

\def\newfam{\alloc@8\fam\chardef\sixt@@n} 

\ifprod@font
\font\fiverm=mtr10 at 5pt
\font\fivebf=mtbx10 at 5pt
\font\fiveit=mtti10 at 5pt
\font\fivesl=mtsl10 at 5pt
\font\fivett=cmtt8 at 5pt     \hyphenchar\fivett=-1
\font\fivecsc=mtcsc10 at 5pt
\font\fivesf=mtss10 at 5pt
\font\fivei=mtmi10 at 5pt      \skewchar\fivei='177
\font\fivesy=mtsy10 at 5pt     \skewchar\fivesy='60

\font\sixrm=mtr10 at 6pt
\font\sixbf=mtbx10 at 6pt
\font\sixit=mtti10 at 6pt
\font\sixsl=mtsl10 at 6pt
\font\sixtt=cmtt8 at 6pt      \hyphenchar\sixtt=-1
\font\sixcsc=mtcsc10 at 6pt
\font\sixsf=mtss10 at 6pt
\font\sixi=mtmi10 at 6pt       \skewchar\sixi='177
\font\sixsy=mtsy10 at 6pt      \skewchar\sixsy='60

\font\sevenrm=mtr10 at 7pt
\font\sevenbf=mtbx10 at 7pt
\font\sevenit=mtti10 at 7pt
\font\sevensl=mtsl10 at 7pt
\font\seventt=cmtt8 at 7pt     \hyphenchar\seventt=-1
\font\sevencsc=mtcsc10 at 7pt
\font\sevensf=mtss10 at 7pt
\font\seveni=mtmi10 at 7pt      \skewchar\seveni='177
\font\sevensy=mtsy10 at 7pt     \skewchar\sevensy='60

\font\eightrm=mtr10 at 8pt
\font\eightbf=mtbx10 at 8pt
\font\eightit=mtti10 at 8pt
\font\eighti=mtmi10 at 8pt      \skewchar\eighti='177
\font\eightsy=mtsy10 at 8pt     \skewchar\eightsy='60
\font\eightsl=mtsl10 at 8pt
\font\eighttt=cmtt8             \hyphenchar\eighttt=-1
\font\eightcsc=mtcsc10 at 8pt
\font\eightsf=mtss10 at 8pt

\font\ninerm=mtr10 at 9pt
\font\ninebf=mtbx10 at 9pt
\font\nineit=mtti10 at 9pt
\font\ninei=mtmi10 at 9pt      \skewchar\ninei='177
\font\ninesy=mtsy10 at 9pt     \skewchar\ninesy='60
\font\ninesl=mtsl10 at 9pt
\font\ninett=cmtt9             \hyphenchar\ninett=-1
\font\ninecsc=mtcsc10 at 9pt
\font\ninesf=mtss10 at 9pt

\font\tenrm=mtr10
\font\tenbf=mtbx10
\font\tenit=mtti10
\font\teni=mtmi10		\skewchar\teni='177
\font\tensy=mtsy10		\skewchar\tensy='60
\font\tenex=cmex10
\font\tensl=mtsl10
\font\tentt=cmtt10		\hyphenchar\tentt=-1
\font\tencsc=mtcsc10
\font\tensf=mtss10

\font\elevenrm=mtr10 at 11pt
\font\elevenbf=mtbx10 at 11pt
\font\elevenit=mtti10 at 11pt
\font\eleveni=mtmi10 at 11pt      \skewchar\eleveni='177
\font\elevensy=mtsy10 at 11pt     \skewchar\elevensy='60
\font\elevensl=mtsl10 at 11pt
\font\eleventt=cmtt10 at 11pt     \hyphenchar\eleventt=-1
\font\elevencsc=mtcsc10 at 11pt
\font\elevensf=mtss10 at 11pt

\font\twelverm=mtr10 at 12pt
\font\twelvebf=mtbx10 at 12pt
\font\twelveit=mtti10 at 12pt
\font\twelvesl=mtsl10 at 12pt
\font\twelvett=cmtt12             \hyphenchar\twelvett=-1
\font\twelvecsc=mtcsc10 at 12pt
\font\twelvesf=mtss10 at 12pt
\font\twelvei=mtmi10 at 12pt      \skewchar\twelvei='177
\font\twelvesy=mtsy10 at 12pt     \skewchar\twelvesy='60

\font\fourteenrm=mtr10 at 14pt
\font\fourteenbf=mtbx10 at 14pt
\font\fourteenit=mtti10 at 14pt
\font\fourteeni=mtmi10 at 14pt      \skewchar\fourteeni='177
\font\fourteensy=mtsy10 at 14pt     \skewchar\fourteensy='60
\font\fourteensl=mtsl10 at 14pt
\font\fourteentt=cmtt12 at 14pt     \hyphenchar\fourteentt=-1
\font\fourteencsc=mtcsc10 at 14pt
\font\fourteensf=mtss10 at 14pt

\font\seventeenrm=mtr10 at 17pt
\font\seventeenbf=mtbx10 at 17pt
\font\seventeenit=mtti10 at 17pt
\font\seventeeni=mtmi10 at 17pt      \skewchar\seventeeni='177
\font\seventeensy=mtsy10 at 17pt     \skewchar\seventeensy='60
\font\seventeensl=mtsl10 at 17pt
\font\seventeentt=cmtt12 at 17pt     \hyphenchar\seventeentt=-1
\font\seventeencsc=mtcsc10 at 17pt
\font\seventeensf=mtss10 at 17pt
\else
\font\fiverm=cmr5
\font\fivei=cmmi5             \skewchar\fivei='177
\font\fivesy=cmsy5            \skewchar\fivesy='60
\font\fivebf=cmbx5

\font\sixrm=cmr6
\font\sixi=cmmi6             \skewchar\sixi='177
\font\sixsy=cmsy6            \skewchar\sixsy='60
\font\sixbf=cmbx6

\font\sevenrm=cmr7
\font\sevenit=cmti7
\font\seveni=cmmi7             \skewchar\seveni='177
\font\sevensy=cmsy7            \skewchar\sevensy='60
\font\sevenbf=cmbx7

\font\eightrm=cmr8
\font\eightbf=cmbx8
\font\eightit=cmti8
\font\eighti=cmmi8			\skewchar\eighti='177
\font\eightsy=cmsy8			\skewchar\eightsy='60
\font\eightsl=cmsl8
\font\eighttt=cmtt8			\hyphenchar\eighttt=-1
\font\eightcsc=cmcsc10 at 8pt
\font\eightsf=cmss8

\font\ninerm=cmr9
\font\ninebf=cmbx9
\font\nineit=cmti9
\font\ninei=cmmi9			\skewchar\ninei='177
\font\ninesy=cmsy9			\skewchar\ninesy='60
\font\ninesl=cmsl9
\font\ninett=cmtt9			\hyphenchar\ninett=-1
\font\ninecsc=cmcsc10 at 9pt
\font\ninesf=cmss9

\font\tenrm=cmr10
\font\tenbf=cmbx10
\font\tenit=cmti10
\font\teni=cmmi10		\skewchar\teni='177
\font\tensy=cmsy10		\skewchar\tensy='60
\font\tenex=cmex10
\font\tensl=cmsl10
\font\tentt=cmtt10		\hyphenchar\tentt=-1
\font\tencsc=cmcsc10
\font\tensf=cmss10

\font\elevenrm=cmr10 scaled \magstephalf
\font\elevenbf=cmbx10 scaled \magstephalf
\font\elevenit=cmti10 scaled \magstephalf
\font\eleveni=cmmi10 scaled \magstephalf	\skewchar\eleveni='177
\font\elevensy=cmsy10 scaled \magstephalf	\skewchar\elevensy='60
\font\elevensl=cmsl10 scaled \magstephalf
\font\eleventt=cmtt10 scaled \magstephalf	\hyphenchar\eleventt=-1
\font\elevencsc=cmcsc10 scaled \magstephalf
\font\elevensf=cmss10 scaled \magstephalf

\font\twelverm=cmr10 scaled \magstep1
\font\twelvebf=cmbx10 scaled \magstep1
\font\twelvei=cmmi10 scaled \magstep1      \skewchar\twelvei='177
\font\twelvesy=cmsy10 scaled \magstep1     \skewchar\twelvesy='60

\font\fourteenrm=cmr10 scaled \magstep2
\font\fourteenbf=cmbx10 scaled \magstep2
\font\fourteenit=cmti10 scaled \magstep2
\font\fourteeni=cmmi10 scaled \magstep2		\skewchar\fourteeni='177
\font\fourteensy=cmsy10 scaled \magstep2	\skewchar\fourteensy='60
\font\fourteensl=cmsl10 scaled \magstep2
\font\fourteentt=cmtt10 scaled \magstep2	\hyphenchar\fourteentt=-1
\font\fourteencsc=cmcsc10 scaled \magstep2
\font\fourteensf=cmss10 scaled \magstep2

\font\seventeenrm=cmr10 scaled \magstep3
\font\seventeenbf=cmbx10 scaled \magstep3
\font\seventeenit=cmti10 scaled \magstep3
\font\seventeeni=cmmi10 scaled \magstep3	\skewchar\seventeeni='177
\font\seventeensy=cmsy10 scaled \magstep3	\skewchar\seventeensy='60
\font\seventeensl=cmsl10 scaled \magstep3
\font\seventeentt=cmtt10 scaled \magstep3	\hyphenchar\seventeentt=-1
\font\seventeencsc=cmcsc10 scaled \magstep3
\font\seventeensf=cmss10 scaled \magstep3
\fi

\def\hexnumber#1{\ifcase#1 0\or1\or2\or3\or4\or5\or6\or7\or8\or9\or
  A\or B\or C\or D\or E\or F\fi}

\def\makestrut{%
  \setbox\strutbox=\hbox{%
    \vrule height.7\baselineskip depth.3\baselineskip width \z@}%
}

\def\baselinestretch{1}
\newskip\tmp@bls

\def\b@ls#1{
  \tmp@bls=#1\relax
  \baselineskip=#1\relax\makestrut
  \normalbaselineskip=\baselinestretch\tmp@bls
  \normalbaselines
}

\def\nostb@ls#1{
  \normalbaselineskip=#1\relax
  \normalbaselines
  \makestrut
}

%

\newfam\scfam  
\newfam\sffam  

\def\mit{\fam\@ne}
\def\cal{\fam\tw@}
\def\em{\ifdim\fontdimen1\font>\z@ \rm\else\it\fi}

\textfont3=\tenex
\scriptfont3=\tenex
\scriptscriptfont3=\tenex

\setbox0=\hbox{\tenex B} \p@renwd=\wd0 

\def\eightpoint{
  \def\rm{\fam0\eightrm}%
  \textfont0=\eightrm \scriptfont0=\sixrm \scriptscriptfont0=\fiverm%
  \textfont1=\eighti  \scriptfont1=\sixi  \scriptscriptfont1=\fivei%
  \textfont2=\eightsy \scriptfont2=\sixsy \scriptscriptfont2=\fivesy%
  \textfont\itfam=\eightit\def\it{\fam\itfam\eightit}%
  \ifprod@font
    \scriptfont\itfam=\sixit
      \scriptscriptfont\itfam=\fiveit
  \else
    \scriptfont\itfam=\eightit
      \scriptscriptfont\itfam=\eightit
  \fi
  \textfont\bffam=\eightbf%
    \scriptfont\bffam=\sixbf%
      \scriptscriptfont\bffam=\fivebf%
  \def\bf{\fam\bffam\eightbf}%
  \textfont\slfam=\eightsl\def\sl{\fam\slfam\eightsl}%
  \ifprod@font
    \scriptfont\slfam=\sixsl
      \scriptscriptfont\slfam=\fivesl
  \else
    \scriptfont\slfam=\eightsl
      \scriptscriptfont\slfam=\eightsl
  \fi
  \textfont\ttfam=\eighttt\def\tt{\fam\ttfam\eighttt}%
  \ifprod@font
    \scriptfont\ttfam=\sixtt
      \scriptscriptfont\ttfam=\fivett
  \else
    \scriptfont\ttfam=\eighttt
      \scriptscriptfont\ttfam=\eighttt
  \fi
  \textfont\scfam=\eightcsc\def\sc{\fam\scfam\eightcsc}%
  \ifprod@font
    \scriptfont\scfam=\sixcsc
      \scriptscriptfont\scfam=\fivecsc
  \else
    \scriptfont\scfam=\eightcsc
      \scriptscriptfont\scfam=\eightcsc
  \fi
  \textfont\sffam=\eightsf\def\sf{\fam\sffam\eightsf}%
  \ifprod@font
    \scriptfont\sffam=\sixsf
      \scriptscriptfont\sffam=\fivesf
  \else
    \scriptfont\sffam=\eightsf
      \scriptscriptfont\sffam=\eightsf
  \fi
  \def\oldstyle{\fam\@ne\eighti}%
  \b@ls{10pt}\rm\@viiipt%
}
\def\@viiipt{}

\def\ninepoint{
  \def\rm{\fam0\ninerm}%
  \textfont0=\ninerm \scriptfont0=\sixrm \scriptscriptfont0=\fiverm%
  \textfont1=\ninei  \scriptfont1=\sixi  \scriptscriptfont1=\fivei%
  \textfont2=\ninesy \scriptfont2=\sixsy \scriptscriptfont2=\fivesy%
  \textfont\itfam=\nineit\def\it{\fam\itfam\nineit}%
  \ifprod@font
    \scriptfont\itfam=\sixit
      \scriptscriptfont\itfam=\fiveit
  \else
    \scriptfont\itfam=\nineit
      \scriptscriptfont\itfam=\nineit
  \fi
  \textfont\bffam=\ninebf%
    \scriptfont\bffam=\sixbf%
      \scriptscriptfont\bffam=\fivebf%
  \def\bf{\fam\bffam\ninebf}%
  \textfont\slfam=\ninesl\def\sl{\fam\slfam\ninesl}%
  \ifprod@font
    \scriptfont\slfam=\sixsl
      \scriptscriptfont\slfam=\fivesl
  \else
    \scriptfont\slfam=\ninesl
      \scriptscriptfont\slfam=\ninesl
  \fi
  \textfont\ttfam=\ninett\def\tt{\fam\ttfam\ninett}%
  \ifprod@font
    \scriptfont\ttfam=\sixtt
      \scriptscriptfont\ttfam=\fivett
  \else
    \scriptfont\ttfam=\ninett
      \scriptscriptfont\ttfam=\ninett
  \fi
  \textfont\scfam=\ninecsc\def\sc{\fam\scfam\ninecsc}%
  \ifprod@font
    \scriptfont\scfam=\sixcsc
      \scriptscriptfont\scfam=\fivecsc
  \else
    \scriptfont\scfam=\ninecsc
      \scriptscriptfont\scfam=\ninecsc
  \fi
  \textfont\sffam=\ninesf\def\sf{\fam\sffam\ninesf}%
  \ifprod@font
    \scriptfont\sffam=\sixsf
      \scriptscriptfont\sffam=\fivesf
  \else
    \scriptfont\sffam=\ninesf
      \scriptscriptfont\sffam=\ninesf
  \fi
  \def\oldstyle{\fam\@ne\ninei}%
  \b@ls{\TextLeading plus \Feathering}\rm\@ixpt%
}
\def\@ixpt{}

\def\tenpoint{
  \def\rm{\fam0\tenrm}%
  \textfont0=\tenrm \scriptfont0=\sevenrm \scriptscriptfont0=\fiverm%
  \textfont1=\teni  \scriptfont1=\seveni  \scriptscriptfont1=\fivei%
  \textfont2=\tensy \scriptfont2=\sevensy \scriptscriptfont2=\fivesy%
  \textfont\itfam=\tenit\def\it{\fam\itfam\tenit}%
  \ifprod@font
    \scriptfont\itfam=\sevenit
      \scriptscriptfont\itfam=\fiveit
  \else
    \scriptfont\itfam=\tenit
      \scriptscriptfont\itfam=\tenit
  \fi
  \textfont\bffam=\tenbf%
    \scriptfont\bffam=\sevenbf%
      \scriptscriptfont\bffam=\fivebf%
  \def\bf{\fam\bffam\tenbf}%
  \textfont\slfam=\tensl\def\sl{\fam\slfam\tensl}%
  \ifprod@font
    \scriptfont\slfam=\sevensl
      \scriptscriptfont\slfam=\fivesl
  \else
    \scriptfont\slfam=\tensl
      \scriptscriptfont\slfam=\tensl
  \fi
  \textfont\ttfam=\tentt\def\tt{\fam\ttfam\tentt}%
  \ifprod@font
    \scriptfont\ttfam=\seventt
      \scriptscriptfont\ttfam=\fivett
  \else
    \scriptfont\ttfam=\tentt
      \scriptscriptfont\ttfam=\tentt
  \fi
  \textfont\scfam=\tencsc\def\sc{\fam\scfam\tencsc}%
  \ifprod@font
    \scriptfont\scfam=\sevencsc
      \scriptscriptfont\scfam=\fivecsc
  \else
    \scriptfont\scfam=\tencsc
      \scriptscriptfont\scfam=\tencsc
  \fi
  \textfont\sffam=\tensf\def\sf{\fam\sffam\tensf}%
  \ifprod@font
    \scriptfont\sffam=\sevensf
      \scriptscriptfont\sffam=\fivesf
  \else
    \scriptfont\sffam=\tensf
      \scriptscriptfont\sffam=\tensf
  \fi
  \def\oldstyle{\fam\@ne\teni}%
  \b@ls{11pt}\rm\@xpt%
}
\def\@xpt{}

\def\elevenpoint{
  \def\rm{\fam0\elevenrm}%
  \textfont0=\elevenrm \scriptfont0=\eightrm \scriptscriptfont0=\sixrm%
  \textfont1=\eleveni  \scriptfont1=\eighti  \scriptscriptfont1=\sixi%
  \textfont2=\elevensy \scriptfont2=\eightsy \scriptscriptfont2=\sixsy%
  \textfont\itfam=\elevenit\def\it{\fam\itfam\elevenit}%
  \ifprod@font
    \scriptfont\itfam=\eightit
      \scriptscriptfont\itfam=\sixit
  \else
    \scriptfont\itfam=\elevenit
      \scriptscriptfont\itfam=\elevenit
  \fi
  \textfont\bffam=\elevenbf%
    \scriptfont\bffam=\eightbf%
      \scriptscriptfont\bffam=\sixbf%
  \def\bf{\fam\bffam\elevenbf}%
  \textfont\slfam=\elevensl\def\sl{\fam\slfam\elevensl}%
  \ifprod@font
    \scriptfont\slfam=\eightsl
      \scriptscriptfont\slfam=\sixsl
  \else
    \scriptfont\slfam=\elevensl
      \scriptscriptfont\slfam=\elevensl
  \fi
  \textfont\ttfam=\eleventt\def\tt{\fam\ttfam\eleventt}%
  \ifprod@font
    \scriptfont\ttfam=\eighttt
      \scriptscriptfont\ttfam=\sixtt
  \else
    \scriptfont\ttfam=\eleventt
      \scriptscriptfont\ttfam=\eleventt
  \fi
  \textfont\scfam=\elevencsc\def\sc{\fam\scfam\elevencsc}%
  \ifprod@font
    \scriptfont\scfam=\eightcsc
      \scriptscriptfont\scfam=\sixcsc
  \else
    \scriptfont\scfam=\elevencsc
      \scriptscriptfont\scfam=\elevencsc
  \fi
  \textfont\sffam=\elevensf\def\sf{\fam\sffam\elevensf}%
  \ifprod@font
    \scriptfont\sffam=\eightsf
      \scriptscriptfont\sffam=\sixsf
  \else
    \scriptfont\sffam=\elevensf
      \scriptscriptfont\sffam=\elevensf
  \fi
  \def\oldstyle{\fam\@ne\eleveni}%
  \b@ls{13pt}\rm\@xipt%
}
\def\@xipt{}

\def\fourteenpoint{
  \def\rm{\fam0\fourteenrm}%
  \textfont0\fourteenrm  \scriptfont0\tenrm  \scriptscriptfont0\sevenrm%
  \textfont1\fourteeni   \scriptfont1\teni   \scriptscriptfont1\seveni%
  \textfont2\fourteensy  \scriptfont2\tensy  \scriptscriptfont2\sevensy%
  \textfont\itfam=\fourteenit\def\it{\fam\itfam\fourteenit}%
  \ifprod@font
    \scriptfont\itfam=\tenit
      \scriptscriptfont\itfam=\sevenit
  \else
    \scriptfont\itfam=\fourteenit
      \scriptscriptfont\itfam=\fourteenit
  \fi
  \textfont\bffam=\fourteenbf%
    \scriptfont\bffam=\tenbf%
      \scriptscriptfont\bffam=\sevenbf%
  \def\bf{\fam\bffam\fourteenbf}%
  \textfont\slfam=\fourteensl\def\sl{\fam\slfam\fourteensl}%
  \ifprod@font
    \scriptfont\slfam=\tensl
      \scriptscriptfont\slfam=\sevensl
  \else
    \scriptfont\slfam=\fourteensl
      \scriptscriptfont\slfam=\fourteensl
  \fi
  \textfont\ttfam=\fourteentt\def\tt{\fam\ttfam\fourteentt}%
  \ifprod@font
    \scriptfont\ttfam=\tentt
      \scriptscriptfont\ttfam=\seventt
  \else
    \scriptfont\ttfam=\fourteentt
      \scriptscriptfont\ttfam=\fourteentt
  \fi
  \textfont\scfam=\fourteencsc\def\sc{\fam\scfam\fourteencsc}%
  \ifprod@font
    \scriptfont\scfam=\tencsc
      \scriptscriptfont\scfam=\sevencsc
  \else
    \scriptfont\scfam=\fourteencsc
      \scriptscriptfont\scfam=\fourteencsc
  \fi
  \textfont\sffam=\fourteensf\def\sf{\fam\sffam\fourteensf}%
  \ifprod@font
    \scriptfont\sffam=\tensf
      \scriptscriptfont\sffam=\sevensf
  \else
    \scriptfont\sffam=\fourteensf
      \scriptscriptfont\sffam=\fourteensf
  \fi
  \def\oldstyle{\fam\@ne\fourteeni}%
  \b@ls{17pt}\rm\@xivpt%
}
\def\@xivpt{}

\def\seventeenpoint{
  \def\rm{\fam0\seventeenrm}%
  \textfont0\seventeenrm  \scriptfont0\twelverm  \scriptscriptfont0\tenrm%
  \textfont1\seventeeni   \scriptfont1\twelvei   \scriptscriptfont1\teni%
  \textfont2\seventeensy  \scriptfont2\twelvesy  \scriptscriptfont2\tensy%
  \textfont\itfam=\seventeenit\def\it{\fam\itfam\seventeenit}%
  \ifprod@font
    \scriptfont\itfam=\twelveit
      \scriptscriptfont\itfam=\tenit
  \else
    \scriptfont\itfam=\seventeenit
      \scriptscriptfont\itfam=\seventeenit
  \fi
  \textfont\bffam=\seventeenbf%
    \scriptfont\bffam=\twelvebf%
      \scriptscriptfont\bffam=\tenbf%
  \def\bf{\fam\bffam\seventeenbf}%
  \textfont\slfam=\seventeensl\def\sl{\fam\slfam\seventeensl}%
  \ifprod@font
    \scriptfont\slfam=\twelvesl
      \scriptscriptfont\slfam=\tensl
  \else
    \scriptfont\slfam=\seventeensl
      \scriptscriptfont\slfam=\seventeensl
  \fi
  \textfont\ttfam=\seventeentt\def\tt{\fam\ttfam\seventeentt}%
  \ifprod@font
    \scriptfont\ttfam=\twelvett
      \scriptscriptfont\ttfam=\tentt
  \else
    \scriptfont\ttfam=\seventeentt
      \scriptscriptfont\ttfam=\seventeentt
  \fi
  \textfont\scfam=\seventeencsc\def\sc{\fam\scfam\seventeencsc}%
  \ifprod@font
    \scriptfont\scfam=\twelvecsc
      \scriptscriptfont\scfam=\tencsc
  \else
    \scriptfont\scfam=\seventeencsc
      \scriptscriptfont\scfam=\seventeencsc
  \fi
  \textfont\sffam=\seventeensf\def\sf{\fam\sffam\seventeensf}%
  \ifprod@font
    \scriptfont\sffam=\twelvesf
      \scriptscriptfont\sffam=\tensf
  \else
    \scriptfont\sffam=\seventeensf
      \scriptscriptfont\sffam=\seventeensf
  \fi
  \def\oldstyle{\fam\@ne\seventeeni}%
  \b@ls{20pt}\rm\@xviipt%
}
\def\@xviipt{}

\lineskip=1pt      \normallineskip=\lineskip
\lineskiplimit=\z@ \normallineskiplimit=\lineskiplimit


\def\,{\relax\ifmmode \mskip\thinmuskip\else \thinspace\fi}
\let\protect=\relax

\long\def\@ifundefined#1#2#3{\expandafter\ifx\csname
  #1\endcsname\relax#2\else#3\fi}




\newtoks\math@groups \math@groups={}
\def\addtom@thgroup#1#2{#1\expandafter{\the#1#2}} 



\def\addtosizeh@ok#1#2#3#4{%
  \expandafter\def\csname @#1pt\endcsname{%
    \def\s@ze{#2}\def\ss@ze{#3}\def\sss@ze{#4}\the\math@groups%
  }%
}



\let\resetsizehook=\addtosizeh@ok


\ifprod@font
  \addtosizeh@ok{viii} {8} {6}  {5}
  \addtosizeh@ok{ix}   {9} {6}  {5}
  \addtosizeh@ok{x}    {10}{7}  {5}
  \addtosizeh@ok{xi}   {11}{8}  {6}
  \addtosizeh@ok{xiv}  {14}{10} {7}
  \addtosizeh@ok{xvii} {17}{12}{10}
\else
  \addtosizeh@ok{viii} {8}     {6}     {5}
  \addtosizeh@ok{ix}   {9}     {6}     {5}
  \addtosizeh@ok{x}    {10}    {7}     {5}
  \addtosizeh@ok{xi}   {10.95} {8}     {6}
  \addtosizeh@ok{xiv}  {14.4}  {10}    {7}
  \addtosizeh@ok{xvii} {17.28} {12}    {10}
\fi

\def\get@font#1#2#3{%
  \edef\fonts@ze{\romannumeral#3}
  \edef\fontn@me{\fonts@ze#1}
  \@ifundefined{\fontn@me}%
    {
     \global\expandafter\font\csname \fontn@me\endcsname=#2 at #3pt}%
    {}%
}

\def\ass@tfont#1#2{%
  \xdef\fam@name{\csname #1\endcsname}%
  \xdef\font@name{\csname #2\endcsname}%
  \let\textfont@name\font@name
  \textfont\fam@name\textfont@name
}

\def\ass@sfont#1#2{%
  \xdef\fam@name{\csname #1\endcsname}%
  \xdef\font@name{\csname #2\endcsname}%
  \let\textfont@name\font@name
  \scriptfont\fam@name\textfont@name
}

\def\ass@ssfont#1#2{%
  \xdef\fam@name{\csname #1\endcsname}%
  \xdef\font@name{\csname #2\endcsname}%
  \let\textfont@name\font@name
  \scriptscriptfont\fam@name\textfont@name
}


\def\NewSymbolFont#1#2{%
  \expandafter\ifx\csname sym#1fam\endcsname\relax 
    \expandafter\newfam\csname sym#1fam\endcsname
    \expandafter\edef\csname sym#1fam\endcsname{\the\allocationnumber}%
    \addtom@thgroup\math@groups{%
      \get@font{#1}{#2}{\s@ze}%
      \ass@tfont{sym#1fam}{\fontn@me}%
      \get@font{#1}{#2}{\ss@ze}%
      \ass@sfont{sym#1fam}{\fontn@me}%
      \get@font{#1}{#2}{\sss@ze}%
      \ass@ssfont{sym#1fam}{\fontn@me}%
    }%
  \else
    \errmessage{Family `#1' already defined}%
  \fi
}


\def\NewMathSymbol#1#2#3#4{%
  \edef\f@mly{\expandafter\hexnumber{\csname sym#3fam\endcsname}}%
  \mathchardef#1="#2\f@mly#4\relax
}


\newif\ifd@f

\def\NewMathDelimiter#1#2#3#4#5#6{%
  \d@ftrue
  \expandafter\ifx\csname sym#3fam\endcsname\relax
    \d@ffalse \errmessage{Family `#3' is not defined}%
  \fi
  \expandafter\ifx\csname sym#5fam\endcsname\relax
    \d@ffalse \errmessage{Family `#5' is not defined}%
  \fi
  \ifd@f
    \edef\f@mly{\expandafter\hexnumber{\csname sym#3fam\endcsname}}%
    \edef\f@mlytw@{\expandafter\hexnumber{\csname sym#5fam\endcsname}}%
    \xdef#1{\delimiter"#2\f@mly #4\f@mlytw@ #6\relax}%
  \fi
}


\def\setboxz@h{\setbox\z@\hbox}
\def\wdz@{\wd\z@}
\def\boxz@{\box\z@}
\def\setbox@ne{\setbox\@ne}
\def\wd@ne{\wd\@ne}

\def\math@atom#1#2{%
   \binrel@{#1}\binrel@@{#2}}
\def\binrel@#1{\setboxz@h{\thinmuskip0mu
  \medmuskip\m@ne mu\thickmuskip\@ne mu$#1\m@th$}%
 \setbox@ne\hbox{\thinmuskip0mu\medmuskip\m@ne mu\thickmuskip
  \@ne mu${}#1{}\m@th$}%
 \setbox\tw@\hbox{\hskip\wd@ne\hskip-\wdz@}}
\def\binrel@@#1{\ifdim\wd2<\z@\mathbin{#1}\else\ifdim\wd\tw@>\z@
 \mathrel{#1}\else{#1}\fi\fi}

\def\m@thit{1}

\def\set@skchar#1{\global\expandafter\skewchar
  \csname\fontn@me\endcsname=#1\relax}

\def\NewMathAlphabet#1#2#3{%
  \def\tst{#3}%
  \ifx\tst\empty\else 
    \expandafter\gdef\csname #1@sc\endcsname{}
  \fi
  \expandafter\def\csname #1\endcsname{
    \protect\csname @#1\endcsname}%
  \expandafter\def\csname @#1\endcsname##1{
    {%
    \begingroup
      \get@font{#1}{#2}{\s@ze}%
      \@ifundefined{#1@sc}{}{\set@skchar{#3}}%
      \ass@tfont{m@thit}{\fontn@me}%
      \get@font{#1}{#2}{\ss@ze}%
      \@ifundefined{#1@sc}{}{\set@skchar{#3}}%
      \ass@sfont{m@thit}{\fontn@me}%
      \get@font{#1}{#2}{\sss@ze}%
      \@ifundefined{#1@sc}{}{\set@skchar{#3}}%
      \ass@ssfont{m@thit}{\fontn@me}%
      \math@atom{##1}{%
      \mathchoice%
        {\hbox{$\m@th\displaystyle##1$}}%
        {\hbox{$\m@th\textstyle##1$}}%
        {\hbox{$\m@th\scriptstyle##1$}}%
        {\hbox{$\m@th\scriptscriptstyle##1$}}}%
    \endgroup
    }%
  }%
}


\newif\iffirstta  \firsttatrue

\def\set@hchar#1{\global\expandafter\hyphenchar
  \csname\fontn@me\endcsname=#1\relax}

\def\NewTextAlphabet#1#2#3{%
  \iffirstta
    \global\firsttafalse
    \newfam\scratchfam
    \edef\scrt@fam{\the\allocationnumber}%
  \fi
  \def\tst{#3}%
  \ifx\tst\empty\else 
    \expandafter\gdef\csname #1@hc\endcsname{}
  \fi
  \expandafter\def\csname #1\endcsname{
    \protect\csname t@#1\endcsname}%
  \long\expandafter\def\csname t@#1\endcsname##1{
    \ifmmode
      \typeout{Warning: do not use \expandafter\string\csname #1\endcsname
        \space in math mode}\fi%
    {%
      \get@font{#1}{#2}{\s@ze}\let\t@xtfnt=\fontn@me\relax
      \@ifundefined{#1@hc}{}{\set@hchar{#3}}%
      \ass@tfont{scrt@fam}{\fontn@me}%
      \get@font{#1}{#2}{\ss@ze}%
      \@ifundefined{#1@hc}{}{\set@hchar{#3}}%
      \ass@sfont{scrt@fam}{\fontn@me}%
      \get@font{#1}{#2}{\sss@ze}%
      \@ifundefined{#1@hc}{}{\set@hchar{#3}}%
      \ass@ssfont{scrt@fam}{\fontn@me}%
      \fam\scratchfam\csname\t@xtfnt\endcsname
    ##1%
    }%
  }%
  \expandafter\def\csname #1shape
    \endcsname{\protect\csname @#1shape\endcsname}%
  \expandafter\def\csname @#1shape\endcsname{
    \ifmmode
      \typeout{Warning: do not use \expandafter\string\csname
        #1shape\endcsname \space in math mode}\fi
      \get@font{#1}{#2}{\s@ze}\let\t@xtfnt=\fontn@me\relax
      \@ifundefined{#1@hc}{}{\set@hchar{#3}}%
      \ass@tfont{scrt@fam}{\fontn@me}%
      \get@font{#1}{#2}{\ss@ze}%
      \@ifundefined{#1@hc}{}{\set@hchar{#3}}%
      \ass@sfont{scrt@fam}{\fontn@me}%
      \get@font{#1}{#2}{\sss@ze}%
      \@ifundefined{#1@hc}{}{\set@hchar{#3}}%
      \ass@ssfont{scrt@fam}{\fontn@me}%
      \fam\scratchfam\csname\t@xtfnt\endcsname
  }%
}


\ifprod@font
  \def\math@itfnt{mtmib10}
  \def\math@syfnt{mtbsy10}
\else
  \def\math@itfnt{cmmib10}
  \def\math@syfnt{cmbsy10}
\fi

\def\m@thsy{2}

\def\bmath{\protect\@bmath}
\def\@bmath#1{%
  {%
  \begingroup
    \get@font{mthit}{\math@itfnt}{\s@ze}\set@skchar{'177}%
    \ass@tfont{m@thit}{\fontn@me}%
    \get@font{mthit}{\math@itfnt}{\ss@ze}\set@skchar{'177}%
    \ass@sfont{m@thit}{\fontn@me}%
    \get@font{mthit}{\math@itfnt}{\sss@ze}\set@skchar{'177}%
    \ass@ssfont{m@thit}{\fontn@me}%
    \get@font{mthsy}{\math@syfnt}{\s@ze}\set@skchar{'60}%
    \ass@tfont{m@thsy}{\fontn@me}%
    \get@font{mthsy}{\math@syfnt}{\ss@ze}\set@skchar{'60}%
    \ass@sfont{m@thsy}{\fontn@me}%
    \get@font{mthsy}{\math@syfnt}{\sss@ze}\set@skchar{'60}%
    \ass@ssfont{m@thsy}{\fontn@me}%
    \math@atom{#1}{%
    \mathchoice%
      {\hbox{$\m@th\displaystyle#1$}}%
      {\hbox{$\m@th\textstyle#1$}}%
      {\hbox{$\m@th\scriptstyle#1$}}%
      {\hbox{$\m@th\scriptscriptstyle#1$}}}%
  \endgroup
  }%
}



\def\diameter{{\ifmmode\mathchoice
{\ooalign{\hfil\hbox{$\displaystyle/$}\hfil\crcr
{\hbox{$\displaystyle\mathchar"20D$}}}}
{\ooalign{\hfil\hbox{$\textstyle/$}\hfil\crcr
{\hbox{$\textstyle\mathchar"20D$}}}}
{\ooalign{\hfil\hbox{$\scriptstyle/$}\hfil\crcr
{\hbox{$\scriptstyle\mathchar"20D$}}}}
{\ooalign{\hfil\hbox{$\scriptscriptstyle/$}\hfil\crcr
{\hbox{$\scriptscriptstyle\mathchar"20D$}}}}
\else{\ooalign{\hfil/\hfil\crcr\mathhexbox20D}}%
\fi}}

\def\sq{\ifmmode\squareforqed\else{\unskip\nobreak\hfil
\penalty50\hskip1em\null\nobreak\hfil\squareforqed
\parfillskip=0pt\finalhyphendemerits=0\endgraf}\fi}
\def\squareforqed{\hbox{\rlap{$\sqcap$}$\sqcup$}}


\def\bbbc{{\mathchoice {\setbox0=\hbox{$\displaystyle\rm C$}\hbox{\hbox
to0pt{\kern0.4\wd0\vrule height0.9\ht0\hss}\box0}}
{\setbox0=\hbox{$\textstyle\rm C$}\hbox{\hbox
to0pt{\kern0.4\wd0\vrule height0.9\ht0\hss}\box0}}
{\setbox0=\hbox{$\scriptstyle\rm C$}\hbox{\hbox
to0pt{\kern0.4\wd0\vrule height0.9\ht0\hss}\box0}}
{\setbox0=\hbox{$\scriptscriptstyle\rm C$}\hbox{\hbox
to0pt{\kern0.4\wd0\vrule height0.9\ht0\hss}\box0}}}}
\def\bbbq{{\mathchoice {\setbox0=\hbox{$\displaystyle\rm
Q$}\hbox{\raise
0.15\ht0\hbox to0pt{\kern0.4\wd0\vrule height0.8\ht0\hss}\box0}}
{\setbox0=\hbox{$\textstyle\rm Q$}\hbox{\raise
0.15\ht0\hbox to0pt{\kern0.4\wd0\vrule height0.8\ht0\hss}\box0}}
{\setbox0=\hbox{$\scriptstyle\rm Q$}\hbox{\raise
0.15\ht0\hbox to0pt{\kern0.4\wd0\vrule height0.7\ht0\hss}\box0}}
{\setbox0=\hbox{$\scriptscriptstyle\rm Q$}\hbox{\raise
0.15\ht0\hbox to0pt{\kern0.4\wd0\vrule height0.7\ht0\hss}\box0}}}}
\def\bbbt{{\mathchoice {\setbox0=\hbox{$\displaystyle\rm
T$}\hbox{\hbox to0pt{\kern0.3\wd0\vrule height0.9\ht0\hss}\box0}}
{\setbox0=\hbox{$\textstyle\rm T$}\hbox{\hbox
to0pt{\kern0.3\wd0\vrule height0.9\ht0\hss}\box0}}
{\setbox0=\hbox{$\scriptstyle\rm T$}\hbox{\hbox
to0pt{\kern0.3\wd0\vrule height0.9\ht0\hss}\box0}}
{\setbox0=\hbox{$\scriptscriptstyle\rm T$}\hbox{\hbox
to0pt{\kern0.3\wd0\vrule height0.9\ht0\hss}\box0}}}}
\def\bbbs{{\mathchoice
{\setbox0=\hbox{$\displaystyle     \rm S$}\hbox{\raise0.5\ht0\hbox
to0pt{\kern0.35\wd0\vrule height0.45\ht0\hss}\hbox
to0pt{\kern0.55\wd0\vrule height0.5\ht0\hss}\box0}}
{\setbox0=\hbox{$\textstyle        \rm S$}\hbox{\raise0.5\ht0\hbox
to0pt{\kern0.35\wd0\vrule height0.45\ht0\hss}\hbox
to0pt{\kern0.55\wd0\vrule height0.5\ht0\hss}\box0}}
{\setbox0=\hbox{$\scriptstyle      \rm S$}\hbox{\raise0.5\ht0\hbox
to0pt{\kern0.35\wd0\vrule height0.45\ht0\hss}\raise0.05\ht0\hbox
to0pt{\kern0.5\wd0\vrule height0.45\ht0\hss}\box0}}
{\setbox0=\hbox{$\scriptscriptstyle\rm S$}\hbox{\raise0.5\ht0\hbox
to0pt{\kern0.4\wd0\vrule height0.45\ht0\hss}\raise0.05\ht0\hbox
to0pt{\kern0.55\wd0\vrule height0.45\ht0\hss}\box0}}}}
\def\bbbz{{\mathchoice {\hbox{$\sf\textstyle Z\kern-0.4em Z$}}
{\hbox{$\sf\textstyle Z\kern-0.4em Z$}}
{\hbox{$\sf\scriptstyle Z\kern-0.3em Z$}}
{\hbox{$\sf\scriptscriptstyle Z\kern-0.2em Z$}}}}


\def\Nulle{0} 
\def\Afe{1}   
\def\Hae{2}   
\def\Hbe{3}   
\def\Hce{4}   
\def\Hde{5}   


\newcount\LastMac       \LastMac=\Nulle

\newskip\half      \half=5.5pt plus 1.5pt minus 2.25pt
\newskip\one       \one=11pt plus 3pt minus 5.5pt
\newskip\onehalf   \onehalf=16.5pt plus 5.5pt minus 8.25pt
\newskip\two       \two=22pt plus 5.5pt minus 11pt

\def\Half{\addvspace{\half}}
\def\One{\addvspace{\one}}
\def\OneHalf{\addvspace{\onehalf}}
\def\Two{\addvspace{\two}}

\def\Raggedright{
  \rightskip=\z@ plus \hsize\relax
}

\def\Fullout{
  \rightskip=\z@\relax
}

\def\Hang#1#2{
  \hangindent=#1%
  \hangafter=#2\relax
}


\newif\ifsp@page
\def\pagestyle#1{\csname ps@#1\endcsname}
\def\thispagestyle#1{\global\sp@pagetrue\gdef\sp@type{#1}}

\def\ps@titlepage{%
  \def\@oddhead{\eightpoint\noindent \the\CatchLine
    \ifprod@font\else\qquad Printed\ \today\qquad
      (MN plain \TeX\ macros\ v\@version)\fi \hfil}%
  \let\@evenhead=\@oddhead
  \def\@oddfoot{\eightpoint\copyright\ \@pubyear\ RAS\hfil}%
  \def\@evenfoot{\hfil\eightpoint\noindent\copyright\ \@pubyear\ RAS}%
}

\def\ps@headings{%
  \def\@oddhead{\elevenpoint\it\noindent
    \hfill\the\RightHeader\hskip1.5em\rm\folio}%
  \def\@evenhead{\elevenpoint\noindent
    \folio\hskip1.5em\it\the\LeftHeader\hfill}%
  \def\@oddfoot{\eightpoint\noindent\copyright\ \@pubyear\ RAS,
    MNRAS {\bf \@volume}, \@pagerange\hfil}%
  \def\@evenfoot{\hfil\eightpoint\copyright\ \@pubyear\ RAS,
    MNRAS {\bf \@volume}, \@pagerange}%
}

\def\ps@plate{%
  \def\@oddhead{\eightpoint\noindent\plt@cap\hfil}%
  \def\@evenhead{\eightpoint\noindent\plt@cap\hfil}%
  \def\@oddfoot{\eightpoint\noindent\copyright\ \@pubyear\ RAS,
    MNRAS {\bf \@volume}, \@pagerange\hfil}%
  \def\@evenfoot{\hfil\eightpoint\copyright\ \@pubyear\ RAS,
    MNRAS {\bf \@volume}, \@pagerange}%
}



\def\title#1{
  \bgroup
    \vbox to 8pt{\vss}%
    \seventeenpoint
    \Raggedright
    \noindent \strut{\bf #1}\par
  \egroup
}

\def\author#1{
  \bgroup
    \ifnum\LastMac=\Afe \OneHalf\else \vskip 21pt\fi
    \fourteenpoint
    \Raggedright
    \noindent \strut #1\par
    \vskip 3pt%
  \egroup
}

\def\affiliation#1{
  \bgroup
    \vskip -4pt%
    \eightpoint
    \Raggedright
    \noindent \strut {\it #1}\par
  \egroup
  \LastMac=\Afe\relax
}

\def\acceptedline#1{
  \bgroup
    \Two
    \eightpoint
    \Raggedright
    \noindent \strut #1\par
  \egroup
}

\long\def\abstract#1{%
  \bgroup
    \vskip 20pt%
    \leftskip 11pc\rightskip\z@
    \noindent{\ninebf ABSTRACT}\par
    \tenpoint
    \Fullout
    \noindent #1\par
  \egroup
}

\long\def\keywords#1{
  \bgroup
    \Half
    \leftskip 11pc\rightskip\z@
    \tenpoint
    \Fullout
    \noindent\hbox{\bf Key words:}\ #1\par
  \egroup
}


\def\maketitle{%
  \EndOpening
  \ifsinglecol \else \MakePage\fi
}



\def\@nameuse#1{\csname #1\endcsname}
\def\arabic#1{\@arabic{\@nameuse{#1}}}
\def\alph#1{\@alph{\@nameuse{#1}}}
\def\Alph#1{\@Alph{\@nameuse{#1}}}
\def\@arabic#1{\number #1}
\def\@Alph#1{\ifcase#1\or A\or B\or C\or D\else\@Ialph{#1}\fi}
\def\@Ialph#1{\ifcase#1\or \or \or \or \or E\or F\or G\or H\or I\or J\or
   K\or L\or M\or N\or O\or P\or Q\or R\or S\or T\or U\or V\or W\or X\or
   Y\or Z\else\errmessage{Counter out of range}\fi}
\def\@alph#1{\ifcase#1\or a\or b\or c\or d\else\@ialph{#1}\fi}
\def\@ialph#1{\ifcase#1\or \or \or \or \or e\or f\or g\or h\or i\or j\or
   k\or l\or m\or n\or o\or p\or q\or r\or s\or t\or u\or v\or w\or x\or y\or
   z\else\errmessage{Counter out of range}\fi}


\newcount\Eqnno
\newcount\SubEqnno

\def\theeq{\arabic{Eqnno}}
\def\thesubeq{\alph{SubEqnno}}

\def\stepeq{\relax
  \global\SubEqnno \z@
  \global\advance\Eqnno \@ne\relax
  {\rm (\theeq)}%
}

\def\startsubeq{\relax
  \global\SubEqnno \z@
  \global\advance\Eqnno \@ne\relax
  \stepsubeq
}

\def\stepsubeq{\relax
  \global\advance\SubEqnno \@ne\relax
  {\rm (\theeq\thesubeq)}%
}


\newcount\Sec        
\newcount\SecSec
\newcount\SecSecSec

\def\thesection{\arabic{Sec}}
\def\thesubsection{\thesection.\arabic{SecSec}}
\def\thesubsubsection{\thesubsection.\arabic{SecSecSec}}

\Sec=\z@

\def\:{\let\@sptoken= } \:  
\def\:{\@xifnch} \expandafter\def\: {\futurelet\@tempc\@ifnch}

\def\@ifnextchar#1#2#3{%
  \let\@tempMACe #1%
  \def\@tempMACa{#2}%
  \def\@tempMACb{#3}%
  \futurelet \@tempMACc\@ifnch%
}

\def\@ifnch{%
\ifx \@tempMACc \@sptoken%
  \let\@tempMACd\@xifnch%
\else%
  \ifx \@tempMACc \@tempMACe%
    \let\@tempMACd\@tempMACa%
  \else%
    \let\@tempMACd\@tempMACb%
  \fi%
\fi%
\@tempMACd%
}

\def\@ifstar#1#2{\@ifnextchar *{\def\@tempMACa*{#1}\@tempMACa}{#2}}

\newskip\@tempskipb

\def\addvspace#1{%
  \ifvmode\else \endgraf\fi%
  \ifdim\lastskip=\z@%
    \vskip #1\relax%
  \else%
    \@tempskipb#1\relax\@xaddvskip%
  \fi%
}

\def\@xaddvskip{%
  \ifdim\lastskip<\@tempskipb%
    \vskip-\lastskip%
    \vskip\@tempskipb\relax%
  \else%
    \ifdim\@tempskipb<\z@%
      \ifdim\lastskip<\z@ \else%
        \advance\@tempskipb\lastskip%
        \vskip-\lastskip\vskip\@tempskipb%
      \fi%
    \fi%
  \fi%
}

\newskip\@tmpSKIP

\def\addpen#1{%
  \ifvmode
    \if@nobreak
    \else
      \ifdim\lastskip=\z@
        \penalty#1\relax
      \else
        \@tmpSKIP=\lastskip
        \vskip -\lastskip
        \penalty#1\vskip\@tmpSKIP
      \fi
    \fi
  \fi
}

\newcount\@clubpen   \@clubpen=\clubpenalty
\newif\if@nobreak    \@nobreakfalse

\def\@noafterindent{%
  \global\@nobreaktrue
  \everypar{\if@nobreak
              \global\@nobreakfalse
              \clubpenalty \@M
              {\setbox\z@\lastbox}%
              \LastMac=\Nulle\relax%
            \else
              \clubpenalty \@clubpen
              \everypar{}%
            \fi}%
}

\newcount\gds@cbrk   \gds@cbrk=-300

\def\@nohdbrk{\interlinepenalty \@M\relax}

\let\@par=\par
\def\@restorepar{\def\par{\@par}}

\newif\if@endpe   \@endpefalse
 
\def\@doendpe{\@endpetrue \@nobreakfalse \LastMac=\Nulle\relax%
     \def\par{\@restorepar\everypar{}\par\@endpefalse}%
              \everypar{\setbox\z@\lastbox\everypar{}\@endpefalse}%
}

\def\section{\@ifstar{\@ssection}{\@section}}

\def\@section#1{
  \if@nobreak
    \everypar{}%
    \ifnum\LastMac=\Hae \addvspace{\half}\fi
  \else
    \addpen{\gds@cbrk}%
    \addvspace{\two}%
  \fi
  \bgroup
    \ninepoint\bf
    \Raggedright
    \global\advance\Sec \@ne
    \ifappendix
      \global\Eqnno=\z@ \global\SubEqnno=\z@\relax
      \def\ch@ck{#1}%
      \ifx\ch@ck\empty \def\c@lon{}\else\def\c@lon{:}\fi
      \noindent\@nohdbrk APPENDIX\ \thesection\c@lon\hskip 0.5em%
        \uppercase{#1}\par
    \else
      \noindent\@nohdbrk\thesection\hskip 1pc \uppercase{#1}\par
    \fi
    \global\SecSec=\z@
  \egroup
  \nobreak
  \vskip\half
  \nobreak
  \@noafterindent
  \LastMac=\Hae\relax
}

\def\@ssection#1{
  \if@nobreak
    \everypar{}%
    \ifnum\LastMac=\Hae \addvspace{\half}\fi
  \else
    \addpen{\gds@cbrk}%
    \addvspace{\two}%
  \fi
  \bgroup
    \ninepoint\bf
    \Raggedright
    \noindent\@nohdbrk\uppercase{#1}\par
  \egroup
  \nobreak
  \vskip\half
  \nobreak
  \@noafterindent
  \LastMac=\Hae\relax
}

\def\subsection{\@ifstar{\@ssubsection}{\@subsection}}

\def\@subsection#1{
  \if@nobreak
    \everypar{}%
    \ifnum\LastMac=\Hae \addvspace{1pt plus 1pt minus .5pt}\fi
  \else
    \addpen{\gds@cbrk}%
    \addvspace{\onehalf}%
  \fi
  \bgroup
    \ninepoint\bf
    \Raggedright
    \global\advance\SecSec \@ne
    \noindent\@nohdbrk\thesubsection \hskip 1pc\relax #1\par
    \global\SecSecSec=\z@
  \egroup
  \nobreak
  \vskip\half
  \nobreak
  \@noafterindent
  \LastMac=\Hbe\relax
}

\def\@ssubsection#1{
  \if@nobreak
    \everypar{}%
    \ifnum\LastMac=\Hae \addvspace{1pt plus 1pt minus .5pt}\fi
  \else
    \addpen{\gds@cbrk}%
    \addvspace{\onehalf}%
  \fi
  \bgroup
    \ninepoint\bf
    \Raggedright
    \noindent\@nohdbrk #1\par
  \egroup
  \nobreak
  \vskip\half
  \nobreak
  \@noafterindent
  \LastMac=\Hbe\relax
}

\def\subsubsection{\@ifstar{\@ssubsubsection}{\@subsubsection}}

\def\@subsubsection#1{
  \if@nobreak
    \everypar{}%
    \ifnum\LastMac=\Hbe \addvspace{1pt plus 1pt minus .5pt}\fi
  \else
    \addpen{\gds@cbrk}%
    \addvspace{\onehalf}%
  \fi
  \bgroup
    \ninepoint\it
    \Raggedright
    \global\advance\SecSecSec \@ne
    \noindent\@nohdbrk\thesubsubsection \hskip 1pc\relax #1\par
  \egroup
  \nobreak
  \vskip\half
  \nobreak
  \@noafterindent
  \LastMac=\Hce\relax
}

\def\@ssubsubsection#1{
  \if@nobreak
    \everypar{}%
    \ifnum\LastMac=\Hbe \addvspace{1pt plus 1pt minus .5pt}\fi
  \else
    \addpen{\gds@cbrk}%
    \addvspace{\onehalf}%
  \fi
  \bgroup
    \ninepoint\it
    \Raggedright
    \noindent\@nohdbrk #1\par
  \egroup
  \nobreak
  \vskip\half
  \nobreak
  \@noafterindent
  \LastMac=\Hce\relax
}

\def\paragraph#1{
  \if@nobreak
    \everypar{}%
  \else
    \addpen{\gds@cbrk}%
    \addvspace{\one}%
  \fi%
  \bgroup%
    \ninepoint\it
    \noindent #1\ \nobreak%
  \egroup
  \LastMac=\Hde\relax
  \ignorespaces
}


\newif\ifappendix

\def\appendix{%
  \global\appendixtrue
  \def\thesection{\Alph{Sec}}%
  \def\thesubsection{\thesection\arabic{SecSec}}%
  \def\theeq{\thesection\arabic{Eqnno}}%
  \Sec=\z@ \SecSec=\z@ \SecSecSec=\z@ \Eqnno=\z@ \SubEqnno=\z@\relax
}




\def\beginlist{%
  \par\if@nobreak \else\addvspace{\half}\fi%
  \bgroup%
    \ninepoint
    \let\item=\list@item%
}

\def\list@item{%
  \par\noindent\hskip 1em\relax%
  \ignorespaces%
}

\def\endlist{\par\egroup\addvspace{\half}\@doendpe}


\def\beginrefs{%
  \par
  \bgroup
    \eightpoint
    \Fullout
    \let\bibitem=\bib@item
}

\def\bib@item{%
  \par\parindent=1.5em\Hang{1.5em}{1}%
  \everypar={\Hang{1.5em}{1}\ignorespaces}%
  \noindent\ignorespaces
}

\def\endrefs{\par\egroup\@doendpe}


\newtoks\CatchLine

\def\@journal{Mon.\ Not.\ R.\ Astron.\ Soc.\ }  
\def\@pubyear{1994}        
\def\@pagerange{000--000}  
\def\@volume{000}          
\def\@microfiche{}         %

\def\pubyear#1{\gdef\@pubyear{#1}\@makecatchline}
\def\pagerange#1{\gdef\@pagerange{#1}\@makecatchline}
\def\volume#1{\gdef\@volume{#1}\@makecatchline}
\def\microfiche#1{\gdef\@microfiche{and Microfiche\ #1}\@makecatchline}

\def\@makecatchline{%
  \global\CatchLine{%
    {\rm \@journal {\bf \@volume},\ \@pagerange\ (\@pubyear)\ \@microfiche}}%
}

\@makecatchline 

\newtoks\LeftHeader

\newtoks\RightHeader

\def\PageHead{
  \begingroup
    \ifsp@page
      \csname ps@\sp@type\endcsname
    \fi
    \ifodd\pageno
      \let\the@head=\@oddhead
    \else
      \let\the@head=\@evenhead
    \fi
    \vbox to \z@{\vskip-22.5\p@%
      \hbox to \PageWidth{\vbox to8.5\p@{}%
        \the@head
      }%
    \vss}%
  \endgroup
  \nointerlineskip
}

\gdef\PageFoot{%
  \nointerlineskip%
  \begingroup
  \ifsp@page
    \csname ps@\sp@type\endcsname
    \global\sp@pagefalse
  \fi
  \vbox to 22pt{\vfil%
    \hbox to \PageWidth{%
      \eightpoint\strut\noindent
      \ifodd\pageno
        \@oddfoot
      \else
        \@evenfoot
      \fi
    }%
  }%
  \endgroup
}

\def\today{%
  \number\day\space
  \ifcase\month\or January\or February\or March\or April\or May\or June\or
    July\or August\or September\or October\or November\or December\fi
  \space\number\year%
}

\def\authorcomment#1{%
  \gdef\PageFoot{%
    \nointerlineskip%
    \vbox to 20pt{\vfil%
      \hbox to \PageWidth{\elevenpoint\noindent \hfil #1 \hfil}}%
  }%
}


\newif\ifplate@page
\newbox\plt@box

\def\beginplatepage{%
  \let\plate=\plate@head
  \let\caption=\fig@caption
  \global\setbox\plt@box=\vbox\bgroup
  \TEMPDIMEN=\PageWidth 
  \hsize=\PageWidth\relax
}

\def\endplatepage{\par\egroup\global\plate@pagetrue}
\def\plate@head#1{\gdef\plt@cap{#1}}


\def\letters{%
  \gdef\folio{\ifnum\pageno<\z@ L\romannumeral-\pageno
    \else L\number\pageno \fi}%
}


\newdimen\mathindent

\global\mathindent=\z@
\global\everydisplay{\global\@dspwd=\displaywidth\displaysetup}


\def\@displaylines#1{
  {}$\displ@y\hbox{\vbox{\halign{$\@lign\hfil\displaystyle##\hfil$\crcr
  #1\crcr}}}${}%
}

\def\@eqalign#1{\null\vcenter{\openup\jot\m@th
  \ialign{\strut\hfil$\displaystyle{##}$&$\displaystyle{{}##}$\hfil
      \crcr#1\crcr}}%
}

\def\@eqalignno#1{
  \global\advance\@dspwd by -\mathindent%
  {}$\displ@y\hbox{\vbox{\halign to\@dspwd%
  {\hfil$\@lign\displaystyle{##}$\tabskip\z@skip
  &$\@lign\displaystyle{{}##}$\hfil\tabskip\centering
  &\llap{$\@lign##$}\tabskip\z@skip\crcr
  #1\crcr}}}${}%
}


\global\let\displaylines=\@displaylines
\global\let\eqalign=\@eqalign
\global\let\eqalignno=\@eqalignno
\global\let\leqalignno=\@eqalignno

\newdimen\@dspwd   \@dspwd=\z@
\newif\if@eqno
\newif\if@leqno
\newtoks\@eqn
\newtoks\@eq

\def\displaysetup#1$${\displaytest#1\eqno\eqno\displaytest}

\def\displaytest#1\eqno#2\eqno#3\displaytest{%
 \if!#3!\ldisplaytest#1\leqno\leqno\ldisplaytest
 \else\@eqnotrue\@leqnofalse\@eqn={#2}\@eq={#1}\fi
 \generaldisplay$$}

\def\ldisplaytest#1\leqno#2\leqno#3\ldisplaytest{%
\@eq={#1}%
 \if!#3!\@eqnofalse\else\@eqnotrue\@leqnotrue
  \@eqn={#2}\fi}

\def\generaldisplay{%
  \if@eqno
    \if@leqno
      \hbox to \displaywidth{\noindent
        \rlap{$\displaystyle\the\@eqn$}%
        \hskip\mathindent$\displaystyle\the\@eq$\hfil}%
    \else
      \hbox to \displaywidth{\noindent
        \hskip\mathindent
        $\displaystyle\the\@eq$\hfil$\displaystyle\the\@eqn$}%
    \fi
  \else
    \hbox to \displaywidth{\noindent
      \hskip\mathindent$\displaystyle\the\@eq$\hfil}%
  \fi
}


\def\@notice{%
  \par\Two%
  \noindent{\b@ls{11pt}\ninerm This paper has been produced using the
    Royal Astronomical Society/Blackwell Science \TeX\ macros.\par}%
}

\outer\def\bye{\@notice\par\vfill\supereject\end}


\def\start@mess{%
  Monthly notices of the RAS journal style (\@typeface)\space
    v\@version,\space \@verdate.%
}

\everyjob{\Warn{\start@mess}}



\newif\if@debug \@debugfalse  

\def\Print#1{\if@debug\immediate\write16{#1}\else \fi}
\def\Warn#1{\immediate\write16{#1}}
\def\wlog#1{}

\newcount\Iteration 

\def\Single{0} \def\Double{1}                 
\def\Figure{0} \def\Table{1}                  

\def\InStack{0}  
\def\InZoneA{1}
\def\InZoneB{2}
\def\InZoneC{3}

\newcount\TEMPCOUNT 
\newdimen\TEMPDIMEN 
\newbox\TEMPBOX     
\newbox\VOIDBOX     

\newcount\LengthOfStack 
\newcount\MaxItems      
\newcount\StackPointer
\newcount\Point         
\newcount\NextFigure    
\newcount\NextTable     
\newcount\NextItem      

\newcount\StatusStack   
\newcount\NumStack      
\newcount\TypeStack     
\newcount\SpanStack     
\newcount\BoxStack      

\newcount\ItemSTATUS    
\newcount\ItemNUMBER    
\newcount\ItemTYPE      
\newcount\ItemSPAN      
\newbox\ItemBOX         
\newdimen\ItemSIZE      

\newdimen\PageHeight    
\newdimen\TextLeading   
\newdimen\Feathering    
\newcount\LinesPerPage  
\newdimen\ColumnWidth   
\newdimen\ColumnGap     
\newdimen\PageWidth     
\newdimen\BodgeHeight   
\newcount\Leading       

\newdimen\ZoneBSize  
\newdimen\TextSize   
\newbox\ZoneABOX     
\newbox\ZoneBBOX     
\newbox\ZoneCBOX     

\newif\ifFirstSingleItem
\newif\ifFirstZoneA
\newif\ifMakePageInComplete
\newif\ifMoreFigures \MoreFiguresfalse 
\newif\ifMoreTables  \MoreTablesfalse  

\newif\ifFigInZoneB 
\newif\ifFigInZoneC 
\newif\ifTabInZoneB 
\newif\ifTabInZoneC

\newif\ifZoneAFullPage

\newbox\MidBOX    
\newbox\LeftBOX
\newbox\RightBOX
\newbox\PageBOX   

\newif\ifLeftCOL  
\LeftCOLtrue

\newdimen\ZoneBAdjust

\newcount\ItemFits
\def\Yes{1}
\def\No{2}


\MaxItems=15
\NextFigure=\z@        
\NextTable=\@ne

\BodgeHeight=6pt
\TextLeading=11pt    
\Leading=11
\Feathering=\z@      
\LinesPerPage=61     
\topskip=\TextLeading
\ColumnWidth=20pc    
\ColumnGap=2pc       

\newskip\ItemSepamount  
\ItemSepamount=\TextLeading plus \TextLeading minus 4pt

\parskip=\z@ plus .1pt
\parindent=18pt
\widowpenalty=\z@
\clubpenalty=10000
\tolerance=1500
\hbadness=1500
\abovedisplayskip=6pt plus 2pt minus 1pt
\belowdisplayskip=6pt plus 2pt minus 1pt
\abovedisplayshortskip=6pt plus 2pt minus 1pt
\belowdisplayshortskip=6pt plus 2pt minus 1pt

\frenchspacing

\ninepoint 

\PageHeight=682pt
\PageWidth=2\ColumnWidth
\advance\PageWidth by \ColumnGap

\pagestyle{headings}




\newcount\DUMMY \StatusStack=\allocationnumber
\newcount\DUMMY \newcount\DUMMY \newcount\DUMMY 
\newcount\DUMMY \newcount\DUMMY \newcount\DUMMY 
\newcount\DUMMY \newcount\DUMMY \newcount\DUMMY
\newcount\DUMMY \newcount\DUMMY \newcount\DUMMY 
\newcount\DUMMY \newcount\DUMMY \newcount\DUMMY

\newcount\DUMMY \NumStack=\allocationnumber
\newcount\DUMMY \newcount\DUMMY \newcount\DUMMY 
\newcount\DUMMY \newcount\DUMMY \newcount\DUMMY 
\newcount\DUMMY \newcount\DUMMY \newcount\DUMMY 
\newcount\DUMMY \newcount\DUMMY \newcount\DUMMY 
\newcount\DUMMY \newcount\DUMMY \newcount\DUMMY

\newcount\DUMMY \TypeStack=\allocationnumber
\newcount\DUMMY \newcount\DUMMY \newcount\DUMMY 
\newcount\DUMMY \newcount\DUMMY \newcount\DUMMY 
\newcount\DUMMY \newcount\DUMMY \newcount\DUMMY 
\newcount\DUMMY \newcount\DUMMY \newcount\DUMMY 
\newcount\DUMMY \newcount\DUMMY \newcount\DUMMY

\newcount\DUMMY \SpanStack=\allocationnumber
\newcount\DUMMY \newcount\DUMMY \newcount\DUMMY 
\newcount\DUMMY \newcount\DUMMY \newcount\DUMMY 
\newcount\DUMMY \newcount\DUMMY \newcount\DUMMY 
\newcount\DUMMY \newcount\DUMMY \newcount\DUMMY 
\newcount\DUMMY \newcount\DUMMY \newcount\DUMMY

\newbox\DUMMY   \BoxStack=\allocationnumber
\newbox\DUMMY   \newbox\DUMMY \newbox\DUMMY 
\newbox\DUMMY   \newbox\DUMMY \newbox\DUMMY 
\newbox\DUMMY   \newbox\DUMMY \newbox\DUMMY 
\newbox\DUMMY   \newbox\DUMMY \newbox\DUMMY 
\newbox\DUMMY   \newbox\DUMMY \newbox\DUMMY

\def\wlog{\immediate\write\m@ne}


\def\GetItemAll#1{%
 \GetItemSTATUS{#1}
 \GetItemNUMBER{#1}
 \GetItemTYPE{#1}
 \GetItemSPAN{#1}
 \GetItemBOX{#1}
}

\def\GetItemSTATUS#1{%
 \Point=\StatusStack
 \advance\Point by #1
 \global\ItemSTATUS=\count\Point
}

\def\GetItemNUMBER#1{%
 \Point=\NumStack
 \advance\Point by #1
 \global\ItemNUMBER=\count\Point
}

\def\GetItemTYPE#1{%
 \Point=\TypeStack
 \advance\Point by #1
 \global\ItemTYPE=\count\Point
}

\def\GetItemSPAN#1{%
 \Point\SpanStack
 \advance\Point by #1
 \global\ItemSPAN=\count\Point
}

\def\GetItemBOX#1{%
 \Point=\BoxStack
 \advance\Point by #1
 \global\setbox\ItemBOX=\vbox{\copy\Point}
 \global\ItemSIZE=\ht\ItemBOX
 \global\advance\ItemSIZE by \dp\ItemBOX
 \TEMPCOUNT=\ItemSIZE
 \divide\TEMPCOUNT by \Leading
 \divide\TEMPCOUNT by 65536
 \advance\TEMPCOUNT \@ne
 \ItemSIZE=\TEMPCOUNT pt
 \global\multiply\ItemSIZE by \Leading
}


\def\JoinStack{%
 \ifnum\LengthOfStack=\MaxItems 
  \Warn{WARNING: Stack is full...some items will be lost!}
 \else
  \Point=\StatusStack
  \advance\Point by \LengthOfStack
  \global\count\Point=\ItemSTATUS
  \Point=\NumStack
  \advance\Point by \LengthOfStack
  \global\count\Point=\ItemNUMBER
  \Point=\TypeStack
  \advance\Point by \LengthOfStack
  \global\count\Point=\ItemTYPE
  \Point\SpanStack
  \advance\Point by \LengthOfStack
  \global\count\Point=\ItemSPAN
  \Point=\BoxStack
  \advance\Point by \LengthOfStack
  \global\setbox\Point=\vbox{\copy\ItemBOX}
  \global\advance\LengthOfStack \@ne
  \ifnum\ItemTYPE=\Figure 
   \global\MoreFigurestrue
  \else
   \global\MoreTablestrue
  \fi
 \fi
}


\def\LeaveStack#1{%
 {\Iteration=#1
 \loop
 \ifnum\Iteration<\LengthOfStack
  \advance\Iteration \@ne
  \GetItemSTATUS{\Iteration}
   \advance\Point by \m@ne
   \global\count\Point=\ItemSTATUS
  \GetItemNUMBER{\Iteration}
   \advance\Point by \m@ne
   \global\count\Point=\ItemNUMBER
  \GetItemTYPE{\Iteration}
   \advance\Point by \m@ne
   \global\count\Point=\ItemTYPE
  \GetItemSPAN{\Iteration}
   \advance\Point by \m@ne
   \global\count\Point=\ItemSPAN
  \GetItemBOX{\Iteration}
   \advance\Point by \m@ne
   \global\setbox\Point=\vbox{\copy\ItemBOX}
 \repeat}
 \global\advance\LengthOfStack by \m@ne
}


\newif\ifStackNotClean

\def\CleanStack{%
 \StackNotCleantrue
 {\Iteration=\z@
  \loop
   \ifStackNotClean
    \GetItemSTATUS{\Iteration}
    \ifnum\ItemSTATUS=\InStack
     \advance\Iteration \@ne
     \else
      \LeaveStack{\Iteration}
    \fi
   \ifnum\LengthOfStack<\Iteration
    \StackNotCleanfalse
   \fi
 \repeat}
}


\def\FindItem#1#2{%
 \global\StackPointer=\m@ne 
 {\Iteration=\z@
  \loop
  \ifnum\Iteration<\LengthOfStack
   \GetItemSTATUS{\Iteration}
   \ifnum\ItemSTATUS=\InStack
    \GetItemTYPE{\Iteration}
    \ifnum\ItemTYPE=#1
     \GetItemNUMBER{\Iteration}
     \ifnum\ItemNUMBER=#2
      \global\StackPointer=\Iteration
      \Iteration=\LengthOfStack 
     \fi
    \fi
   \fi
  \advance\Iteration \@ne
 \repeat}
}


\def\FindNext{%
 \global\StackPointer=\m@ne 
 {\Iteration=\z@
  \loop
  \ifnum\Iteration<\LengthOfStack
   \GetItemSTATUS{\Iteration}
   \ifnum\ItemSTATUS=\InStack
    \GetItemTYPE{\Iteration}
   \ifnum\ItemTYPE=\Figure
    \ifMoreFigures
      \global\NextItem=\Figure
      \global\StackPointer=\Iteration
      \Iteration=\LengthOfStack 
    \fi
   \fi
   \ifnum\ItemTYPE=\Table
    \ifMoreTables
      \global\NextItem=\Table
      \global\StackPointer=\Iteration
      \Iteration=\LengthOfStack 
    \fi
   \fi
  \fi
  \advance\Iteration \@ne
 \repeat}
}


\def\ChangeStatus#1#2{%
 \Point=\StatusStack
 \advance\Point by #1
 \global\count\Point=#2
}



\def\Zone{\InZoneA}

\ZoneBAdjust=\z@

\def\MakePage{
 \global\ZoneBSize=\PageHeight
 \global\TextSize=\ZoneBSize
 \global\ZoneAFullPagefalse
 \global\topskip=\TextLeading
 \MakePageInCompletetrue
 \MoreFigurestrue
 \MoreTablestrue
 \FigInZoneBfalse
 \FigInZoneCfalse
 \TabInZoneBfalse
 \TabInZoneCfalse
 \global\FirstSingleItemtrue
 \global\FirstZoneAtrue
 \global\setbox\ZoneABOX=\box\VOIDBOX
 \global\setbox\ZoneBBOX=\box\VOIDBOX
 \global\setbox\ZoneCBOX=\box\VOIDBOX
 \loop
  \ifMakePageInComplete
 \FindNext
 \ifnum\StackPointer=\m@ne
  \NextItem=\m@ne
  \MoreFiguresfalse
  \MoreTablesfalse
 \fi
 \ifnum\NextItem=\Figure
   \FindItem{\Figure}{\NextFigure}
   \ifnum\StackPointer=\m@ne \global\MoreFiguresfalse
   \else
    \GetItemSPAN{\StackPointer}
    \ifnum\ItemSPAN=\Single \def\Zone{\InZoneB}\relax
     \ifFigInZoneC \global\MoreFiguresfalse\fi
    \else
     \def\Zone{\InZoneA}
     \ifFigInZoneB \def\Zone{\InZoneC}\fi
    \fi
   \fi
   \ifMoreFigures\Print{}\FigureItems\fi
 \fi
\ifnum\NextItem=\Table
   \FindItem{\Table}{\NextTable}
   \ifnum\StackPointer=\m@ne \global\MoreTablesfalse
   \else
    \GetItemSPAN{\StackPointer}
    \ifnum\ItemSPAN=\Single\relax
     \ifTabInZoneC \global\MoreTablesfalse\fi
    \else
     \def\Zone{\InZoneA}
     \ifTabInZoneB \def\Zone{\InZoneC}\fi
    \fi
   \fi
   \ifMoreTables\Print{}\TableItems\fi
 \fi
   \MakePageInCompletefalse 
   \ifMoreFigures\MakePageInCompletetrue\fi
   \ifMoreTables\MakePageInCompletetrue\fi
 \repeat
 \ifZoneAFullPage
  \global\TextSize=\z@
  \global\ZoneBSize=\z@
  \global\vsize=\z@\relax
  \global\topskip=\z@\relax
  \vbox to \z@{\vss}
  \eject
 \else
 \global\advance\ZoneBSize by -\ZoneBAdjust
 \global\vsize=\ZoneBSize
 \global\hsize=\ColumnWidth
 \global\ZoneBAdjust=\z@
 \ifdim\TextSize<23pt
 \Warn{}
 \Warn{* Making column fall short: TextSize=\the\TextSize *}
 \vskip-\lastskip\eject\fi
 \fi
}

\def\MakeRightCol{
 \global\TextSize=\ZoneBSize
 \MakePageInCompletetrue
 \MoreFigurestrue
 \MoreTablestrue
 \global\FirstSingleItemtrue
 \global\setbox\ZoneBBOX=\box\VOIDBOX
 \def\Zone{\InZoneB}
 \loop
  \ifMakePageInComplete
 \FindNext
 \ifnum\StackPointer=\m@ne
  \NextItem=\m@ne
  \MoreFiguresfalse
  \MoreTablesfalse
 \fi
 \ifnum\NextItem=\Figure
   \FindItem{\Figure}{\NextFigure}
   \ifnum\StackPointer=\m@ne \MoreFiguresfalse
   \else
    \GetItemSPAN{\StackPointer}
    \ifnum\ItemSPAN=\Double\relax
     \MoreFiguresfalse\fi
   \fi
   \ifMoreFigures\Print{}\FigureItems\fi
 \fi
 \ifnum\NextItem=\Table
   \FindItem{\Table}{\NextTable}
   \ifnum\StackPointer=\m@ne \MoreTablesfalse
   \else
    \GetItemSPAN{\StackPointer}
    \ifnum\ItemSPAN=\Double\relax
     \MoreTablesfalse\fi
   \fi
   \ifMoreTables\Print{}\TableItems\fi
 \fi
   \MakePageInCompletefalse 
   \ifMoreFigures\MakePageInCompletetrue\fi
   \ifMoreTables\MakePageInCompletetrue\fi
 \repeat
 \ifZoneAFullPage
  \global\TextSize=\z@
  \global\ZoneBSize=\z@
  \global\vsize=\z@\relax
  \global\topskip=\z@\relax
  \vbox to \z@{\vss}
  \eject
 \else
 \global\vsize=\ZoneBSize
 \global\hsize=\ColumnWidth
 \ifdim\TextSize<23pt
 \Warn{}
 \Warn{* Making column fall short: TextSize=\the\TextSize *}
 \vskip-\lastskip\eject\fi
\fi
}

\def\FigureItems{
 \Print{Considering...}
 \ShowItem{\StackPointer}
 \GetItemBOX{\StackPointer} 
 \GetItemSPAN{\StackPointer}
  \CheckFitInZone 
  \ifnum\ItemFits=\Yes
   \ifnum\ItemSPAN=\Single
     \ChangeStatus{\StackPointer}{\InZoneB} 
     \global\FigInZoneBtrue
     \ifFirstSingleItem
      \hbox{}\vskip-\BodgeHeight
     \global\advance\ItemSIZE by \TextLeading
     \fi
     \unvbox\ItemBOX\ItemSep
     \global\FirstSingleItemfalse
     \global\advance\TextSize by -\ItemSIZE
     \global\advance\TextSize by -\TextLeading
   \else
    \ifFirstZoneA
     \global\advance\ItemSIZE by \TextLeading
     \global\FirstZoneAfalse\fi
    \global\advance\TextSize by -\ItemSIZE
    \global\advance\TextSize by -\TextLeading
    \global\advance\ZoneBSize by -\ItemSIZE
    \global\advance\ZoneBSize by -\TextLeading
    \ifFigInZoneB\relax
     \else
     \ifdim\TextSize<3\TextLeading
     \global\ZoneAFullPagetrue
     \fi
    \fi
    \ChangeStatus{\StackPointer}{\Zone}
    \ifnum\Zone=\InZoneC \global\FigInZoneCtrue\fi
  \fi
   \Print{TextSize=\the\TextSize}
   \Print{ZoneBSize=\the\ZoneBSize}
  \global\advance\NextFigure \@ne
   \Print{This figure has been placed.}
  \else
   \Print{No space available for this figure...holding over.}
   \Print{}
   \global\MoreFiguresfalse
  \fi
}

\def\TableItems{
 \Print{Considering...}
 \ShowItem{\StackPointer}
 \GetItemBOX{\StackPointer} 
 \GetItemSPAN{\StackPointer}
  \CheckFitInZone 
  \ifnum\ItemFits=\Yes
   \ifnum\ItemSPAN=\Single
    \ChangeStatus{\StackPointer}{\InZoneB}
     \global\TabInZoneBtrue
     \ifFirstSingleItem
      \hbox{}\vskip-\BodgeHeight
     \global\advance\ItemSIZE by \TextLeading
     \fi
     \unvbox\ItemBOX\ItemSep
     \global\FirstSingleItemfalse
     \global\advance\TextSize by -\ItemSIZE
     \global\advance\TextSize by -\TextLeading
   \else
    \ifFirstZoneA
    \global\advance\ItemSIZE by \TextLeading
    \global\FirstZoneAfalse\fi
    \global\advance\TextSize by -\ItemSIZE
    \global\advance\TextSize by -\TextLeading
    \global\advance\ZoneBSize by -\ItemSIZE
    \global\advance\ZoneBSize by -\TextLeading
    \ifFigInZoneB\relax
     \else
     \ifdim\TextSize<3\TextLeading
     \global\ZoneAFullPagetrue
     \fi
    \fi
    \ChangeStatus{\StackPointer}{\Zone}
    \ifnum\Zone=\InZoneC \global\TabInZoneCtrue\fi
   \fi
  \global\advance\NextTable \@ne
   \Print{This table has been placed.}
  \else
  \Print{No space available for this table...holding over.}
   \Print{}
   \global\MoreTablesfalse
  \fi
}


\def\CheckFitInZone{%
{\advance\TextSize by -\ItemSIZE
 \advance\TextSize by -\TextLeading
 \ifFirstSingleItem
  \advance\TextSize by \TextLeading
 \fi
 \ifnum\Zone=\InZoneA\relax
  \else \advance\TextSize by -\ZoneBAdjust
 \fi
 \ifdim\TextSize<3\TextLeading \global\ItemFits=\No
 \else \global\ItemFits=\Yes\fi}
}

\def\BeginOpening{%
  \ninepoint
  \thispagestyle{titlepage}%
  \global\setbox\ItemBOX=\vbox\bgroup%
    \hsize=\PageWidth%
    \hrule height \z@
    \ifsinglecol\vskip 6pt\fi 
}

\let\begintopmatter=\BeginOpening  

\def\EndOpening{%
  \One
  \egroup
  \ifsinglecol
    \box\ItemBOX%
    \vskip\TextLeading plus 2\TextLeading
    \@noafterindent
  \else
    \ItemNUMBER=\z@%
    \ItemTYPE=\Figure
    \ItemSPAN=\Double
    \ItemSTATUS=\InStack
    \JoinStack
  \fi
}


\newif\if@here  \@herefalse

\def\no@float{\global\@heretrue}
\let\nofloat=\relax 

\def\beginfigure{%
  \@ifstar{\global\@dfloattrue \@bfigure}{\global\@dfloatfalse \@bfigure}%
}

\def\@bfigure#1{%
  \par
  \if@dfloat
    \ItemSPAN=\Double
    \TEMPDIMEN=\PageWidth
  \else
    \ItemSPAN=\Single
    \TEMPDIMEN=\ColumnWidth
  \fi
  \ifsinglecol
    \TEMPDIMEN=\PageWidth
  \else
    \ItemSTATUS=\InStack
    \ItemNUMBER=#1%
    \ItemTYPE=\Figure
  \fi
  \bgroup
    \hsize=\TEMPDIMEN
    \global\setbox\ItemBOX=\vbox\bgroup
      \eightpoint\nostb@ls{10pt}%
      \let\caption=\fig@caption
      \ifsinglecol \let\nofloat=\no@float\fi
}

\def\fig@caption#1{%
  \vskip 5.5pt plus 6pt%
  \bgroup 
    \eightpoint\nostb@ls{10pt}%
    \setbox\TEMPBOX=\hbox{#1}%
    \ifdim\wd\TEMPBOX>\TEMPDIMEN
      \noindent \unhbox\TEMPBOX\par
    \else
      \hbox to \hsize{\hfil\unhbox\TEMPBOX\hfil}%
    \fi
  \egroup
}

\def\endfigure{%
  \par\egroup 
  \egroup
  \ifsinglecol
    \if@here \midinsert\global\@herefalse\else \topinsert\fi
      \unvbox\ItemBOX
    \endinsert
  \else
    \JoinStack
    \Print{Processing source for figure \the\ItemNUMBER}%
  \fi
}


\newbox\tab@cap@box
\def\tab@caption#1{\global\setbox\tab@cap@box=\hbox{#1\par}}

\newtoks\tab@txt@toks
\long\def\tab@txt#1{\global\tab@txt@toks={#1}\global\table@txttrue}

\newif\iftable@txt  \table@txtfalse
\newif\if@dfloat    \@dfloatfalse

\def\begintable{%
  \@ifstar{\global\@dfloattrue \@btable}{\global\@dfloatfalse \@btable}%
}

\def\@btable#1{%
  \par
  \if@dfloat
    \ItemSPAN=\Double
    \TEMPDIMEN=\PageWidth
  \else
    \ItemSPAN=\Single
    \TEMPDIMEN=\ColumnWidth
  \fi
  \ifsinglecol
    \TEMPDIMEN=\PageWidth
  \else
    \ItemSTATUS=\InStack
    \ItemNUMBER=#1%
    \ItemTYPE=\Table
  \fi
  \bgroup
    \eightpoint\nostb@ls{10pt}%
    \global\setbox\ItemBOX=\vbox\bgroup
      \let\caption=\tab@caption
      \let\tabletext=\tab@txt
      \ifsinglecol \let\nofloat=\no@float\fi
}

\def\endtable{%
  \par\egroup 
  \egroup
  \setbox\TEMPBOX=\hbox to \TEMPDIMEN{%
    \eightpoint\nostb@ls{10pt}%
    \hss
    \vbox{%
      \hsize=\wd\ItemBOX
      \ifvoid\tab@cap@box
      \else
        \noindent\unhbox\tab@cap@box
        \vskip 5.5pt plus 6pt%
      \fi
      \box\ItemBOX
      \iftable@txt
        \vskip 10pt%
        \noindent\the\tab@txt@toks
        \global\table@txtfalse
      \fi
    }%
    \hss
  }%
  \ifsinglecol
    \if@here \midinsert\global\@herefalse\else \topinsert\fi
      \box\TEMPBOX
    \endinsert
  \else
    \global\setbox\ItemBOX=\box\TEMPBOX
    \JoinStack
    \Print{Processing source for table \the\ItemNUMBER}%
  \fi
}

\def\UnloadZoneA{%
\FirstZoneAtrue
 \Iteration=\z@
  \loop
   \ifnum\Iteration<\LengthOfStack
    \GetItemSTATUS{\Iteration}
    \ifnum\ItemSTATUS=\InZoneA
     \GetItemBOX{\Iteration}
     \ifFirstZoneA \vbox to \BodgeHeight{\vfil}%
     \FirstZoneAfalse\fi
     \unvbox\ItemBOX\ItemSep
     \LeaveStack{\Iteration}
     \else
     \advance\Iteration \@ne
   \fi
 \repeat
}

\def\UnloadZoneC{%
\Iteration=\z@
  \loop
   \ifnum\Iteration<\LengthOfStack
    \GetItemSTATUS{\Iteration}
    \ifnum\ItemSTATUS=\InZoneC
     \GetItemBOX{\Iteration}
     \ItemSep\unvbox\ItemBOX
     \LeaveStack{\Iteration}
     \else
     \advance\Iteration \@ne
   \fi
 \repeat
}


\def\ShowItem#1{
  {\GetItemAll{#1}
  \Print{\the#1:
  {TYPE=\ifnum\ItemTYPE=\Figure Figure\else Table\fi}
  {NUMBER=\the\ItemNUMBER}
  {SPAN=\ifnum\ItemSPAN=\Single Single\else Double\fi}
  {SIZE=\the\ItemSIZE}}}
}

\def\ShowStack{%
 \Print{}
 \Print{LengthOfStack = \the\LengthOfStack}
 \ifnum\LengthOfStack=\z@ \Print{Stack is empty}\fi
 \Iteration=\z@
 \loop
 \ifnum\Iteration<\LengthOfStack
  \ShowItem{\Iteration}
  \advance\Iteration \@ne
 \repeat
}

\def\B#1#2{%
\hbox{\vrule\kern-0.4pt\vbox to #2{%
\hrule width #1\vfill\hrule}\kern-0.4pt\vrule}
}


\newif\ifsinglecol   \singlecolfalse

\def\onecolumn{%
  \global\output={\singlecoloutput}%
  \global\hsize=\PageWidth
  \global\vsize=\PageHeight
  \global\ColumnWidth=\hsize
  \global\TextLeading=12pt
  \global\Leading=12
  \global\singlecoltrue
  \global\let\onecolumn=\relax
  \global\let\footnote=\sing@footnote
  \global\let\vfootnote=\sing@vfootnote
  \ninepoint 
  \message{(Single column)}%
}

\def\singlecoloutput{%
  \shipout\vbox{\PageHead\vbox to \PageHeight{\pagebody\vss}\PageFoot}%
  \advancepageno
  \ifplate@page
    \shipout\vbox{%
      \sp@pagetrue
      \def\sp@type{plate}%
      \global\plate@pagefalse
      \PageHead\vbox to \PageHeight{\unvbox\plt@box\vfil}\PageFoot%
    }%
    \message{[plate]}%
    \advancepageno
  \fi
  \ifnum\outputpenalty>-\@MM \else\dosupereject\fi%
}

\def\ItemSep{\vskip\ItemSepamount\relax}

\def\ItemSepbreak{\par\ifdim\lastskip<\ItemSepamount
  \removelastskip\penalty-200\ItemSep\fi%
}


\let\@@endinsert=\endinsert 

\def\endinsert{\egroup 
  \if@mid \dimen@\ht\z@ \advance\dimen@\dp\z@ \advance\dimen@12\p@
    \advance\dimen@\pagetotal \advance\dimen@-\pageshrink
    \ifdim\dimen@>\pagegoal\@midfalse\p@gefalse\fi\fi
  \if@mid \ItemSep\box\z@\ItemSepbreak
  \else\insert\topins{\penalty100 
    \splittopskip\z@skip
    \splitmaxdepth\maxdimen \floatingpenalty\z@
    \ifp@ge \dimen@\dp\z@
    \vbox to\vsize{\unvbox\z@\kern-\dimen@}
    \else \box\z@\nobreak\ItemSep\fi}\fi\endgroup%
}


\def\gobbleone#1{}
\def\gobbletwo#1#2{}
\let\footnote=\gobbletwo 
\let\vfootnote=\gobbleone

\def\sing@footnote#1{\let\@sf\empty 
  \ifhmode\edef\@sf{\spacefactor\the\spacefactor}\/\fi
  \hbox{$^{\hbox{\eightpoint #1}}$}\@sf\sing@vfootnote{#1}%
}

\def\sing@vfootnote#1{\insert\footins\bgroup\eightpoint\b@ls{9pt}%
  \interlinepenalty\interfootnotelinepenalty
  \splittopskip\ht\strutbox 
  \splitmaxdepth\dp\strutbox \floatingpenalty\@MM
  \leftskip\z@skip \rightskip\z@skip \spaceskip\z@skip \xspaceskip\z@skip
  \noindent $^{\scriptstyle\hbox{#1}}$\hskip 4pt%
    \footstrut\futurelet\next\fo@t%
}

\def\footnoterule{\kern-3\p@ \hrule height \z@ \kern 3\p@}

\skip\footins=19.5pt plus 12pt minus 1pt
\count\footins=1000
\dimen\footins=\maxdimen

\def\note#1#2{%
  \let\@sf=\empty \ifhmode\edef\@sf{\spacefactor\the\spacefactor}\/\fi
  #1\insert\footins\bgroup
    \eightpoint\b@ls{10pt}\rm
    \interlinepenalty\interfootnotelinepenalty
    \splitmaxdepth\dp\strutbox \floatingpenalty\@MM
    \leftskip\z@skip \rightskip\z@skip \spaceskip\z@skip \xspaceskip\z@skip
    \noindent\footstrut #1$\,$#2\strut\par
  \egroup
  \@sf\relax}


\def\landscape{%
  \global\TEMPDIMEN=\PageWidth
  \global\PageWidth=\PageHeight
  \global\PageHeight=\TEMPDIMEN
  \global\let\landscape=\relax
  \onecolumn
  \message{(landscape)}%
  \raggedbottom
}


\output{%
  \ifLeftCOL
    \global\setbox\LeftBOX=\vbox to \ZoneBSize{\box255\unvbox\ZoneBBOX
      \ifvoid\footins\else
        \vskip\skip\footins\unvbox\footins\fi
    }%
    \global\LeftCOLfalse
    \MakeRightCol
  \else
    \setbox\RightBOX=\vbox to \ZoneBSize{\box255\unvbox\ZoneBBOX
      \ifvoid\footins\else
        \vskip\skip\footins\unvbox\footins\fi
    }%
    \setbox\MidBOX=\hbox{\box\LeftBOX\hskip\ColumnGap\box\RightBOX}%
    \setbox\PageBOX=\vbox to \PageHeight{%
      \UnloadZoneA\box\MidBOX\UnloadZoneC}%
    \shipout\vbox{\PageHead\vbox to \PageHeight{\box\PageBOX\vss}\PageFoot}%
    \advancepageno
    \ifplate@page
      \shipout\vbox{%
        \sp@pagetrue
        \def\sp@type{plate}%
        \global\plate@pagefalse
        \PageHead\vbox to \PageHeight{\unvbox\plt@box\vfil}\PageFoot%
      }%
      \message{[plate]}%
      \advancepageno
    \fi
    \global\LeftCOLtrue
    \CleanStack
    \MakePage
  \fi
}


\Warn{\start@mess}

\newif\ifCUPmtplainloaded 
\ifprod@font
  \global\CUPmtplainloadedtrue
\fi


\catcode `\@=12 



\def\etal{et al.}
\def\jcd{Christensen-Dalsgaard}

\def\eg{{e.g.}}
\def\cf{{cf.}}

\def\etal{{et al.}}
\def\Fsurf{F_{\rm surf}}

\begintopmatter
\title{Effects of errors in the solar radius on helioseismic inferences}
\author{Sarbani Basu }

\affiliation{School of Natural Sciences, Institute for Advanced Study, 
Olden Lane, Princeton NJ 08540, U.S.A.}
\acceptedline{Accepted \ . Received \ }

\abstract{Frequencies of intermediate-degree f-modes of the Sun seem to 
indicate that
the solar radius is smaller than what is normally used in constructing
solar models. We investigate the possible consequences of an error in radius
on results for solar structure obtained using helioseismic inversions.
It is shown that solar sound speed will be overestimated if oscillation
frequencies are inverted using reference models with a larger radius. 
Using solar models with  radius of 695.78 Mm and new data sets,
the base of the solar convection zone is estimated to be at radial distance
of $0.7135\pm 0.0005$ of the solar radius. The helium abundance in the 
convection
zone as determined using models with OPAL equation of state is 
$0.248\pm 0.001$, where the errors reflect the estimated systematic errors
in the calculation, the statistical errors being much smaller.
Assuming that the OPAL opacities used in the construction of the solar
models are correct, the surface $Z/X$ is estimated to be $0.0245\pm 0.0006$.
}

\keywords {Sun: oscillations --- Sun: interior --- Sun: abundances ---
Sun: convection}

\maketitle

\section{Introduction}

Helioseismology has proved to be a remarkable tool that can be used to
study the solar interior. Helioseismic inversion techniques have been used to
probe details of the solar structure (cf., Gough et al.~1996; Basu et al.~1996;
Kosovichev et al.~1997), abundances (cf., Basu \& Antia 1995; Kosovichev 1996;  Antia \& Chitre
1997a), and rotation rate (cf., Thompson et al.~1996).
Most inversions done so far assume that the
radius of the Sun is known precisely and the value used is 695.99 Mm
(Allen 1976).

Recent data from the Global Oscillation Network Group (GONG),
and  from the SOI/MDI instrument on board
the SOHO spacecraft, 
allow the determination of the intermediate-degree solar f-mode frequencies
very precisely.  
It has been shown by Antia (1997) and Schou et al.~(1997) that the 
ratio of  the newly determined solar f-mode frequencies  and frequencies
of the  f-modes of standard solar models is very nearly a constant which
differs significantly from unity. 
Since the f-mode frequencies are largely insensitive to the
detailed structure of the Sun and depend mainly on the global parameters,
the simplest explanation of this difference is that the radius assumed
for the models is not the radius of the Sun. Antia (1997) estimated
the solar radius to be 695.78 Mm using the GONG
data, while Schou et al.~(1997) using MDI data
show that the apparent photospheric solar radius (695.99 Mm) used 
to calibrate the models should be reduced by approximately 0.3 Mm.
This difference between the radius of the Sun and of solar models
used as reference models for helioseismic inversions can lead to errors 
in our estimation of the solar structure.

The exact value of the solar radius as determined by the f-modes depends to
some extent on the surface physics used in the models 
(cf., Antia \& Chitre 1997b).
Although the frequencies of solar f-modes are remarkably insensitive
to the structure of the solar interior, they do depend to a small extent
on the details of the surface layers.  The modes are  affected particularly
by the formulation used to
calculate the convective flux and atmospheric opacities.
Models constructed with the Canuto-Mazzitelli (1991) formulation for
calculating the convective flux
show that a smaller reduction in radius is needed for the 
model and observed f-mode frequencies to match, 
compared with the reduction needed for  models constructed
with the mixing length formalism. 
This difference is most probably 
 responsible for the apparent difference of results found by
Schou et al.~(1997) using 
the data from the SOI/MDI instrument on board the SOHO
spacecraft, and  by Antia (1997) using data  from GONG.
As a result, there is still some uncertainty 
in the estimate of solar radius.
 However, if models with the same physics are
used, then the GONG and MDI data give very similar results (Antia \&
Chitre 1997b).

An inversion for solar structure ({\eg}, Dziembowski, Pamyatnykh \&
Sienkiewicz 1990; D\"appen {\etal} 1991; Antia \& Basu 1994a;
Dziembowski {\etal}~1994) generally proceeds through a linearisation
of the equations of stellar oscillations around a known reference
model.  The differences  between the 
structure of the Sun and the reference model are then related to the 
differences in the frequencies of the Sun and the model by kernels.
Non-adiabatic effects and other errors in modelling the surface layers
give rise to frequency shifts (Cox \& Kidman
1984; Balmforth 1992) which are not accounted for by the variational
principle. In the absence of any reliable formulation, these
effects have been taken into account in an {\it ad hoc} manner by
including an arbitrary function of frequency in the variational
formulation (Dziembowski et al.~1990). 
Thus the fractional change in frequency of a mode can be expressed
in terms of fractional
changes in the structure of the model and  a surface term.

When the oscillation equation is linearised --- under the assumption
of hydrostatic equilibrium --- the fractional change in the
frequency can be related to the fractional changes in two of the
functions that define the structure of the models.  Thus,
$$
\eqalign{ {\delta \omega_i \over \omega_i}&
= \int K_{1,2}^i(r){ \delta f_{1}(r) \over f_1(r)}d r +
 \int K_{2,1}^i(r) {\delta f_{2}(r)\over f_2(r)} d r \cr &\quad
 +{\Fsurf(\omega_i)\over E_i} \; \cr}\eqno\eqname\full
$$
({\cf}, Dziembowski {\etal} 1990).  Here $\delta \omega_i$ is the
difference in the frequency $\omega_i$ of the $i$th mode between the
solar data and a reference model.  The functions $f_{1}$ and $f_{2}$
are an appropriate pair of functions like sound speed and density, or
density and adiabatic index  $\Gamma_1$ etc.  The kernels $K_{1,2}^i$
and $K_{2,1}^i$ are known functions of the reference model which
relate the changes in frequency to the changes in $f_{1}$ and $f_{2}$
respectively; and $E_i$ is the inertia of the mode, normalized by the
photospheric amplitude of the displacement.  The term $\Fsurf$
results from the near-surface errors.

\beginfigure{1}
\hbox to 0 pt{\hskip -1.5cm
\vbox to 6.5 true cm{\vskip -1.75 true cm
\epsfysize=10.50 true cm\epsfbox{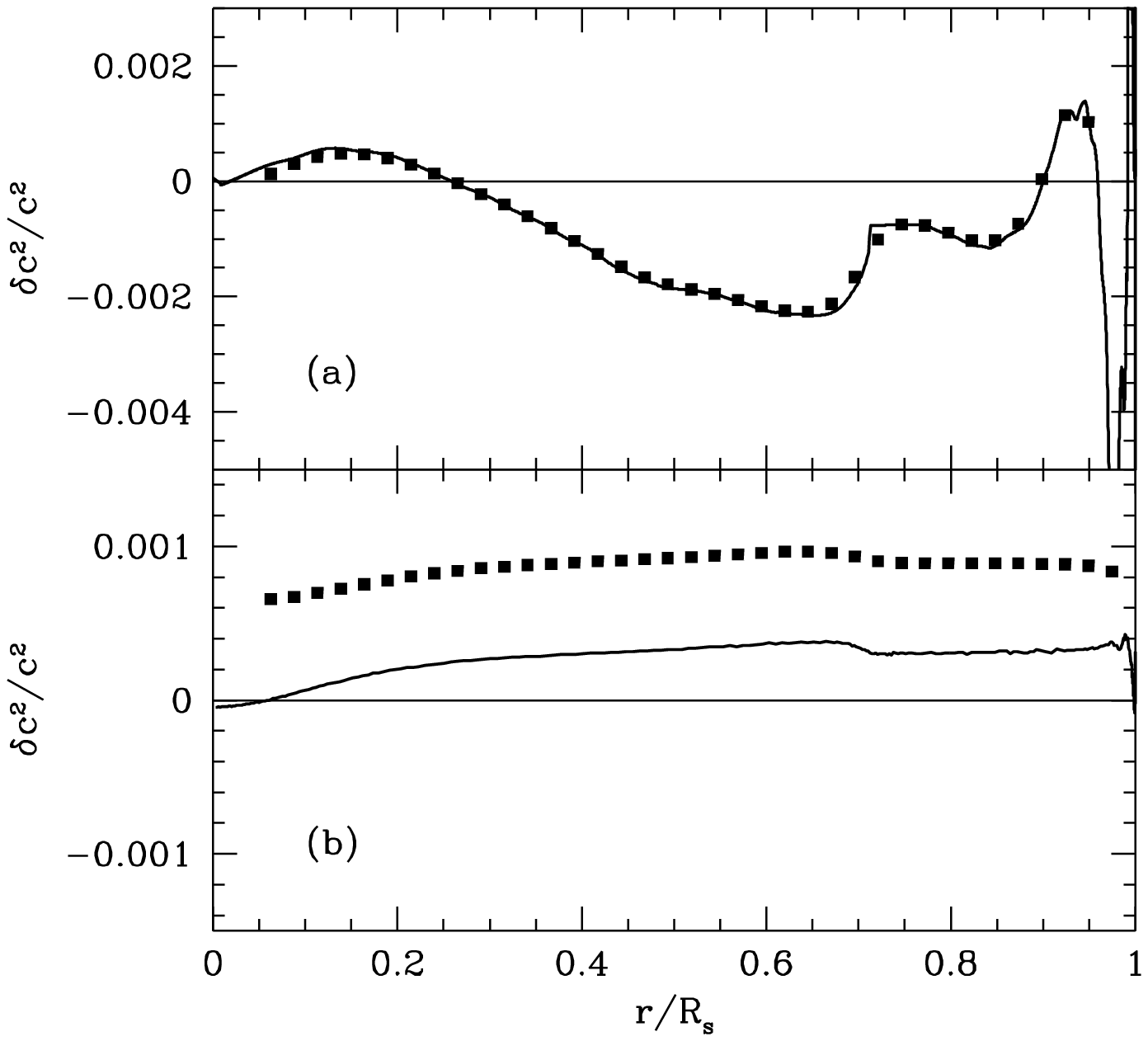}\vskip -2.25 true cm}}
\caption{\bf Fig.~1 \rm The result of sound-speed inversions between 
different models.
The continuous line is the exact sound-speed difference
between the two models, the points are the results obtained by inverting the
frequency differences between the two models. In Panel (a), both the reference
and test models have the same radius, 695.99Mm, the difference
between the two models is the equation of state used. In Panel (b), the 
models have identical input physics, but radii of the
two models are different. The reference model has a radius of 695.99 Mm while
the test model has a radius of 695.78 Mm.
}
\endfigure

The kernels $K_{1,2}^i$ and $K_{2,1}^i$ are normally calculated assuming
that there is no difference between the  radius of the Sun and the
reference model (see Antia \& Basu (1994a) for the steps involved
in calculating the kernels for the pair $(c^2,\rho)$ ). Thus inversions 
of  Eq.~\full\ are likely to introduce systematic errors if the radius of the
reference model is different from that of the Sun or test model.
This can be seen in Fig.~1 for inversion between two models. 
Hence, our results on solar sound speed are also likely
to be modified if the radius of the solar model is different from that of
the Sun, and initial results suggest that this is indeed the case
(Antia 1997).

In this work we show how a mismatch of radius affects solar sound-speed
results. Further, since most of our knowledge of the solar interior,
e.g., depth of the convection zone (CZ), overshoot below the solar
convection zone, helium abundance in the
solar envelope, etc.,
are derived from the sound speed, we re-calibrate these quantities 
in view of possible changes in the value of solar radius.
While determining the  helium abundance we have looked into different sources
of systematic errors, like  opacities and depth of CZ, which had not been
adequately examined in earlier studies. We also attempt to put limits on
the heavy-element abundance in the solar envelope.
Unlike most previous works which utilized oscillation data 
from the Big Bear Solar Observatory,   in this work
more precise data from GONG and SOI/MDI have also been used to
determine the depth of the solar convection zone and the helium
abundance.

The rest of the paper is organized as follows. In Section 2, we discuss the
change in deduced sound-speed and density profiles in the Sun that
results from using models that correspond to the reduced radius as
determined from the f-mode frequencies.  In Sections 3 and
4 we discuss the effect of the radius change on the deduced depth of the
solar convection zone  and helium abundance respectively. In Section 5 
we try to investigate if an error in equation of state can explain the
remaining difference
between the sound speed of the Sun and solar models inside the convection
zone. 
We present the conclusions from this study in Section 6.

\section{Inversion results}

\beginfigure{2}
\hbox to 0 pt{\hskip -1.5cm
\vbox to 7.5 true cm{\vskip -1.75 true cm
\epsfysize=11.50 true cm\epsfbox{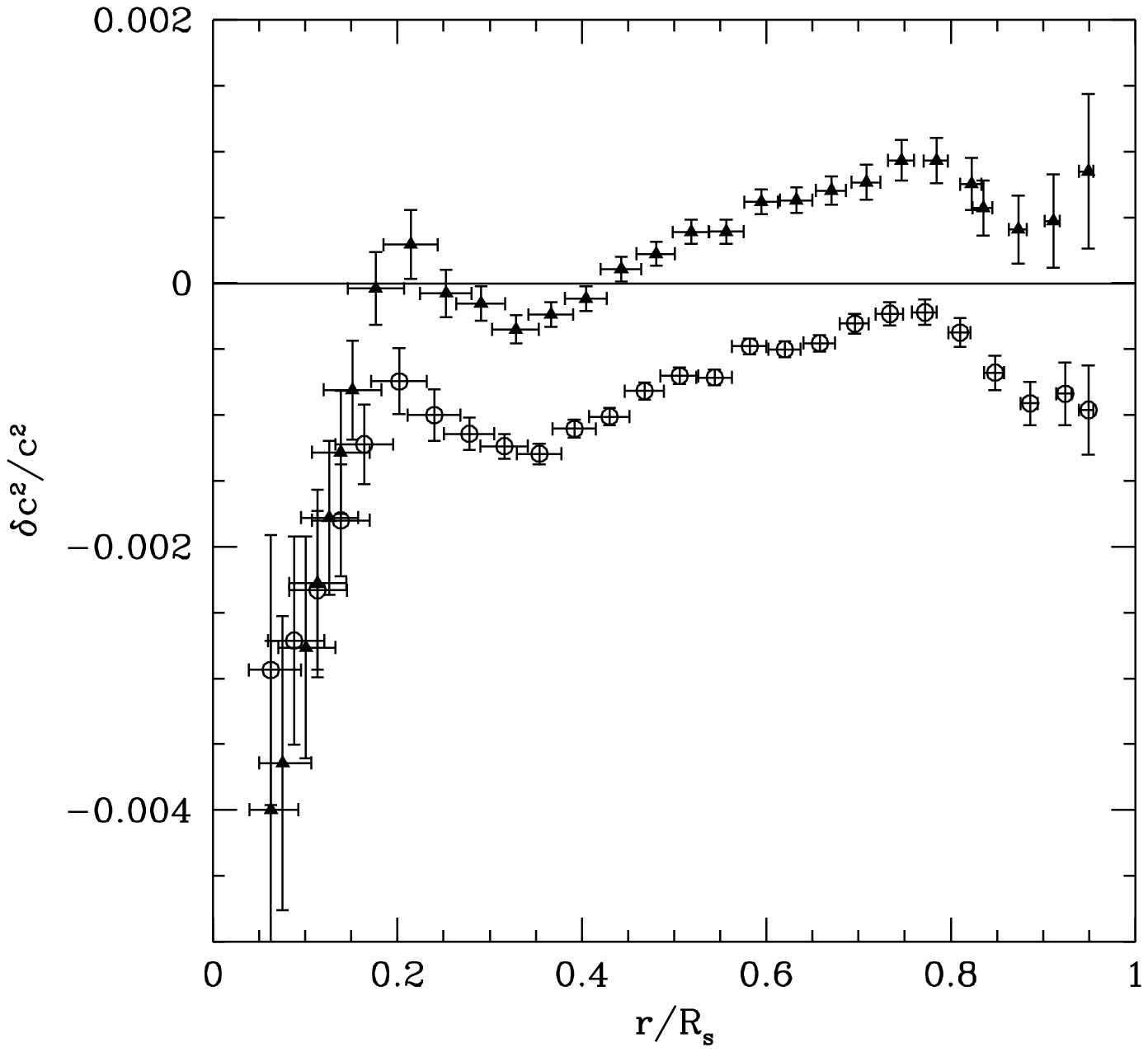}\vskip -2.25 true cm}}
\caption{\bf Fig.~2 \rm The relative sound-speed difference between the
Sun and two reference models R99 and R78 at fixed fractional radius. 
The differences are in the sense $(\hbox{Sun}-\hbox{model})/\hbox{model}$.
The triangles are the results obtained with
a reference model of R99 (radius 695.99 Mm), the circles are obtained 
with reference model R78 (radius 695.78 Mm).
The vertical error bars show the 1$\sigma$ error
in the inversions due to errors in the data. The horizontal error-bars are a
measure of the resolution.
The sound-speed difference between the two models can be seen
in Fig.~1(b).
}
\endfigure

\beginfigure{3}
\hbox to 0 pt{\hskip -0.750cm
\vbox to 9.5 true cm{\vskip -0.5 true cm
\epsfysize=10.25 true cm\epsfbox{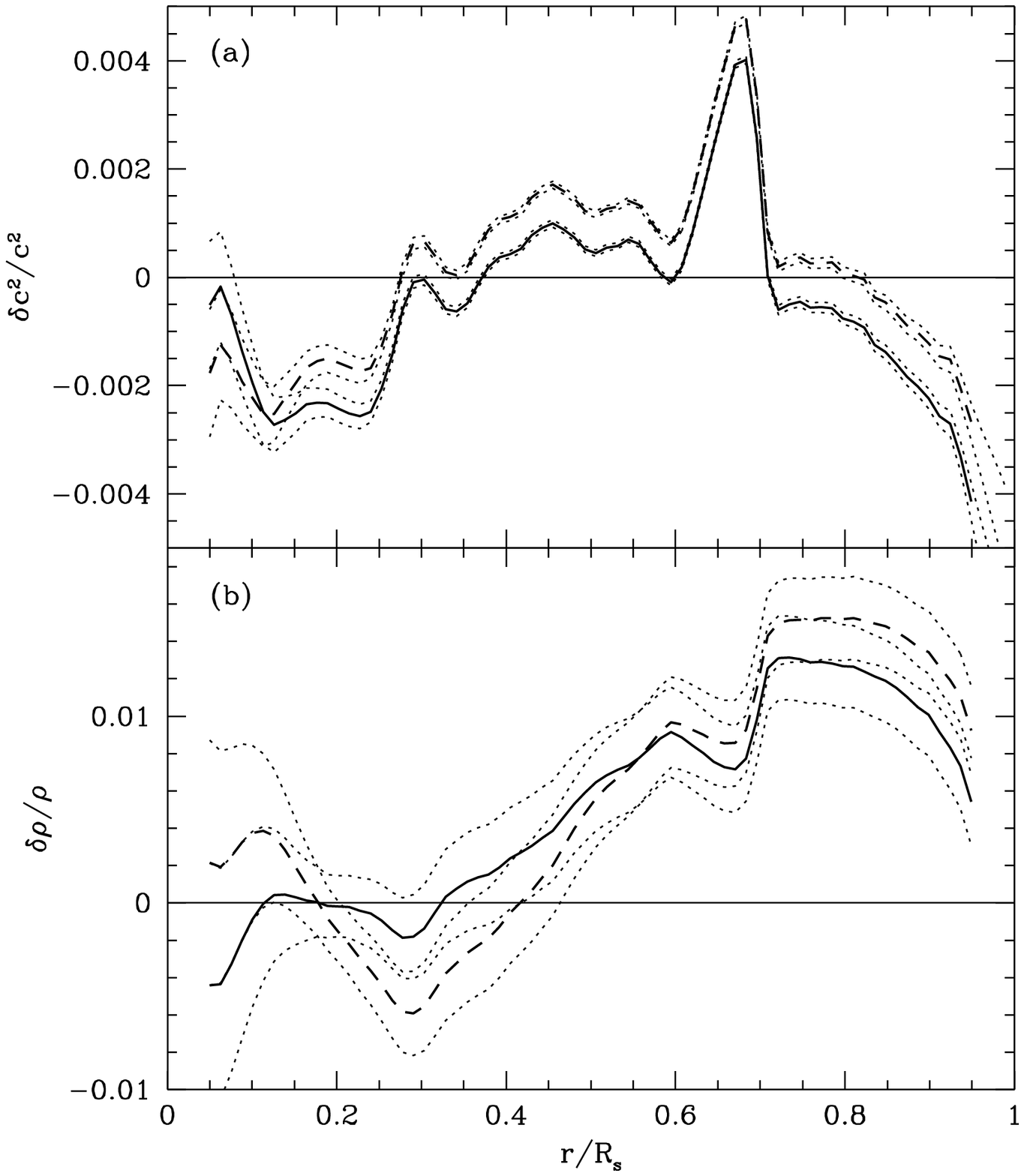}\vskip -0.5 true cm}}
\caption{\bf Fig.~3 \rm (a) The sound-speed and (b) density inside the 
Sun as obtained by the two reference models R99 and R78 shown with 
respect to Model S of \jcd\ et al.~(1996). In both panels,
 the continuous line represents the result
obtained with model R78 ($R_{\rm s}=695.78$ Mm) and the dashed line
that that obtained using model R99 ($R_{\rm s}=695.99$ Mm). The dotted
lines represent the $1\sigma$ error envelope.
}
\endfigure

 In this work we have  used the subtractive optimally localized 
averages (SOLA) method of Pijpers \& Thompson (1992) to invert 
Eq.~\full. The procedure as applied to determine the sound-speed
difference is discussed in Basu et al.~(1996).

Three sets of helioseismic data have been used for this work. The first set is
the
data from the Big Bear Solar Observatory (Libbrecht, Woodard and Kaufman,
1990) combined with the low degree ($0\le \ell \le 3$) modes from the
Birmingham Solar Oscillation Network (Elsworth et al.~1994), this set is 
referred to as the BBSO set. The second set consists of  the frequencies 
obtained from the average spectra of months 4 -- 14 obtained by the Global 
Oscillation Network Group (hereafter referred to as the GONG data), and the
third set are 
the frequencies  from the data obtained by the SOI/MDI instrument 
on board the SOHO spacecraft during its first 144 days in operation
(Rhodes et al.~1997), which we shall refer to as the MDI data.
We restrict ourselves to the frequency range of 1 to 3.5 mHz. The BBSO
set is further restricted to a degree range of $\ell \le 140$ to avoid
frequencies
determined by the ridge fitting method which may have large systematic
errors.

We have used two reference models with identical physical inputs,
but different radii  for the inversions.  Both models are static-models,
constructed with the OPAL equation of state (Rogers, Swenson \& Iglesias 1996)
 and OPAL opacities 
(Iglesias \& Rogers 1996) which are supplemented by the low 
temperature opacities
of Kurucz (1991).  The CM formulation is used to calculate the convective
flux. To construct the
model we use the helioseismically determined hydrogen abundance profile from
Antia \&  Chitre (1997a), and also use the heavy elements abundance
profile adopted
by them. The surface ratio of heavy elements to hydrogen abundance was
constrained
to be 0.0245, the value determined from observations by Grevesse \& Noels
 (1993).
The radii used are the standard value of 695.99 Mm (Model R99) and the
reduced  value 695.78 Mm (Model R78) as determined by Antia (1997).
In this paper  we denote the radius of the solar models
as $R_{\rm s}$, to avoid confusion with $R_\odot$, which is normally
assumed to have the fixed value of 695.99 Mm.

The sound-speed difference between the Sun and the two models are shown in
Fig.~2. The sound-speed difference between the two models has 
been shown earlier in Fig.~1(b), and hence it is easy to see that there is
a real difference between solar sound-speed deduced by using the
two models.  The same is true for solar density.
To make comparisons easier,  we have represented the calculated
sound-speed profiles as differences with respect to the Model S of 
Christensen-Dalsgaard et al.~(1996) in Fig.~3(a).

Thus it can be seen that there is a significant difference between the
results obtained using the two models, which arises because of the
difference in radius. In fact, for sound-speed, the
relative difference between the two solutions is of the same order as
the relative sound-speed difference between the Sun and the two reference 
models. Thus it is important to invert solar oscillation frequencies using
reference models with the same radius as the Sun. Fig.~3(b) shows the
density results
obtained using the two models. Again there is a real difference in the
result obtained, though in the case of density, the changes are within
the $1\sigma$ error limits.

The sound-speed difference near the surface
between Model S and the Sun shows a very sharp dip which may be attributed
to the difference in radius. In fact, the sound-speed
difference at fixed radius between models R99 and R78 shows a sharp
decrease at the surface,
which may lead one to assume that the change in the radius of the reference
model will reduce the dip.
However, this decrease is not very evident in 
differences taken at fixed fractional radius (cf., Fig.~1b), which is
what is used in the inversions, and hence it is
unlikely
to affect the inversion results significantly.
In fact, we can see from Figs.~2 and 3
that the surface behaviour of the solutions is not very different regardless
of the radius of the model used for inverting the solar oscillation
frequencies.
Thus we believe that the sharp decline in the relative sound-speed difference
is more a result  of the other physical properties of the model.

\beginfigure{4}
\hbox to 0 pt{\hskip -1.5cm
\vbox to 9.5 true cm{\vskip -0.5 true cm
\epsfysize=10.50 true cm\epsfbox{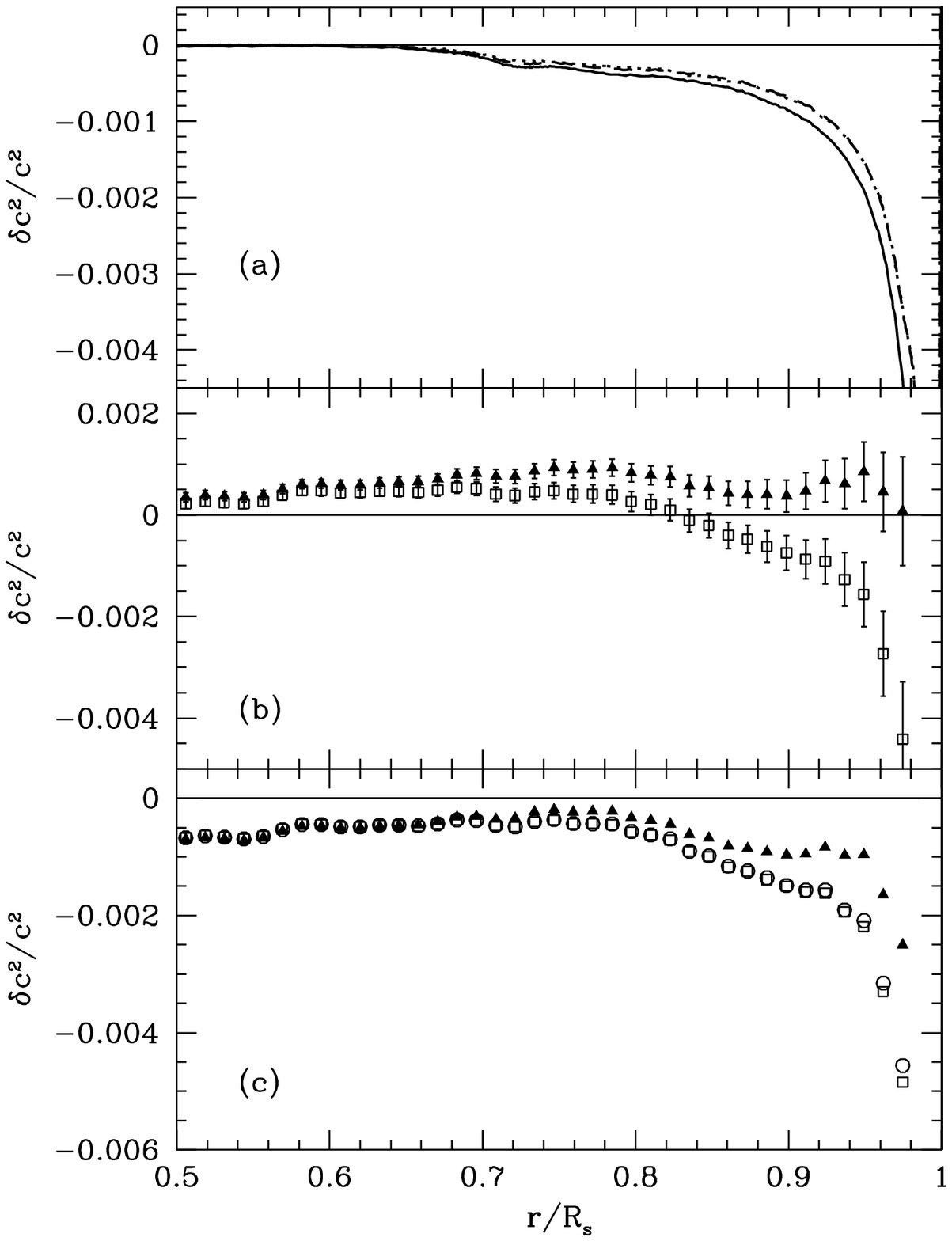}\vskip -0.5 true cm}}
\caption{\bf Fig.~4 \rm (a) The relative sound-speed difference between the
reference models R99 and R78 and the models with modified physics. The 
differences are calculated in the sense
($\hbox{reference}-\hbox{modified}$).
The continuous line is the difference between models R78 and OP78, the
dotted line is the difference between models R99 and MLT99 and the 
dashed line is the difference between models R78 and MLT78. Note that the
dotted and dashed lines more or less coincide.
(b) The relative sound-speed difference
between the Sun and model R99 (triangles) and MLT99 (squares).
(c) The relative sound-speed difference between the Sun and models R78 
(triangles),
MLT78 (squares) and OP78 (circles). The error-bars are omitted in this 
panel for the sake of clarity.
}
\endfigure

To confirm this point we have constructed reference solar models using
physical inputs which give rise to large changes in the surface
layers as compared to the two reference models used above. One set consists 
of models with the two different radii and constructed using the mixing length
theory (MLT) for calculating the convective flux instead of CM formulation.
The models are
called MLT99 (for $R_{\rm s}=695.99$ Mm) and MLT78 (for $R_{\rm s}=695.78$ Mm).
We also have  one low radius model where instead of  Kurucz opacities, the OPAL
opacities as used as far as the table continue. Below that opacities 
from Alexander (1975) are used. However, in that region the temperature 
stratification is obtained using the $T-\tau$ relation of
Vernazza et al.~(1981). This model is called OP78.

The relative difference in sound-speed between our original reference
models and the modified models are shown in Fig.~4(a). Note that the modified
models have a higher sound speed in the surface layers as compared to the 
models with the CM formulation of convection and Kurucz opacities,
 thus the sound-speed difference between the Sun and the modified models
is likely to show a dip at the surface.
The results of the inversion  are shown in Fig.~4 (b) and (c). The new models,
irrespective of radius, do show a sharp decrease in the sound-speed difference
in the surface layers. Thus it is clear that a difference in radius is
not the cause of the near-surface plunge in the relative sound-speed
difference seen between the Sun and Model S.

Fig.~4 also supports the result of Basu \& Antia (1994a) that the models
constructed with the CM formulation are closer to the Sun than those with
MLT formulation. It also appears that the Kurucz (1991) opacities 
fit solar data better than just the OPAL opacities supplemented
by those from Alexander (1975). This appears to support the results of
Antia \& Basu (1997) using just the frequency differences.
 However,  there remains a small dip near the surface which points 
to  remaining deficiencies in model R78.

The change in the deduced solar sound-speed profiles because of changes in
the radius necessitates the re-calibration of quantities derived from the
sound-speed difference. We therefore, look next at the depth and helium
abundance of the solar convection zone.

\section{Depth of the convection zone}

The transition of the temperature gradient from the adiabatic to radiative
values at the base of the solar
convection zone (CZ)  leaves its signature on the sound speed.
Thus helioseismic measurements of the sound speed enables a determination 
of the position of the base of the convection zone (cf.,
\jcd, Gough \& Thompson 1991; Kosovichev \& Fedorova 1991; 
Guzik \& Cox 1993;
Basu \& Antia  1997).

\subsection{The Procedure}
   
We use the technique described in Basu \& Antia (1997) to determine the position
of the CZ base, $r_b$.
To recapitulate briefly, if there are two otherwise similar
solar models with different depths of the convection zone, then 
just below the base of the convection zone
the model with a deeper convection zone will have a larger
sound speed than the other.
This observable difference of sound speeds can be calibrated to find the
convection-zone depth of a test model or the Sun.

Asymptotically, the frequency differences between a solar model
and the Sun, or between two solar models can be written as (\jcd,
Gough \& Thompson 1989)
$$
S(w){\delta \omega\over \omega}=H_1(w)+H_2(\omega),\eqno\eqname\dif$$
where
$w=\omega/(\ell+0.5)$, and  $S(w)$ is a known function for a given
reference model.
The functions $H_1(w)$ and $H_2(\omega)$ can be found by a
least-squares fit to the known frequency differences.
$H_1(w)$ 
 can be inverted to obtain the sound-speed difference, $\delta c/c$,
 between the reference model and the Sun.
 However, that is not required as $H_1(w)$ itself
can be used to determine the convection-zone depth.

If $\phi(w)$ is the $H_1(w)$ between two solar models which differ
only in the depth of the convection zone,
then $H_1(w)$ for any other pair of models can be written as
$$
H_1(w)=\beta\phi(w)+H_s(w),\eqno\eqname\calib$$
where $H_s(w)$ is a smooth component of $H_1(w)$ which
results from sound-speed differences that arise from differences
in the equation of state, abundances, surface physics etc.,
and  the first term is the contribution to $H_1(w)$ due to the
sound-speed difference caused by the difference in $r_b$, the
position of the base of the CZ. Thus if $\beta$ is
determined by a least squares fit, the unknown $r_b$ of the Sun can
be obtained.

We determine $\beta$ for a series of calibration models with
different $r_b$,
and interpolate to find the position for which $\beta=0$. 
The error in the CZ depth arising from those in the observed frequencies is 
determined by Monte Carlo simulations.

As is clear from the above discussion, we need to use models with a specified
depth of CZ. This is easier  with models of the solar envelope,
where position of the CZ base and the helium abundance of the solar
envelope can be specified. 
As Basu \& Antia (1997) have shown, the largest source of systematic
error in the results is the hydrogen abundance profile. This is because the 
sound
speed near the base of the CZ is affected not only by the change in the
temperature gradient, but also by the change in the mean molecular
weight due to gravitational settling of helium. Since the $X$ and $Z$ profiles
are not known exactly, we use models with three types of profiles:
{\parindent=25 pt

\item{ND:} The hydrogen and heavy element abundance profiles from
the no-diffusion model of Bahcall \& Pinsonneault (1992).

\item{DIF:} The hydrogen abundance profile from solar model of
Bahcall \& Pinsonneault (1992) incorporating the diffusion of
helium and $Z$ profile from Proffitt (1994).

\item{INV:} The hydrogen abundance profile obtained from helioseismic 
inversions (Antia \& Chitre 1997a) and the  $Z$ profile adopted in that
paper.

}

\begintable{1}
\caption{\bf Table 1. \rm CZ results for the test models}
{\tablet{8.4 true cm}{#\hfil&&\hfil $#$\cr
\tabmidrule
Test model & \multispan{2}{\hfil Deduced $r_b/R_{\rm s}$\hfil }\cr
           &    R_{\rm s}=695.99\hbox{Mm } & R_{\rm s}=695.78\hbox{Mm} \cr
\tabmidrule
         & \multispan{2}{Calib. models with INV $X$ profile}\cr
\tabmidrule
INV99-0.712 & 0.712000& 0.712258\cr
INV78-0.712  & 0.711746& 0.712000\cr
INV57-0.712 & 0.711498& 0.711746\cr
\tabmidrule
} }
\endtable

\beginfigure{5}
\hbox to 0 pt{\hskip -1.5cm
\vbox to 6.5 true cm{\vskip -1.75 true cm
\epsfysize=10.50 true cm\epsfbox{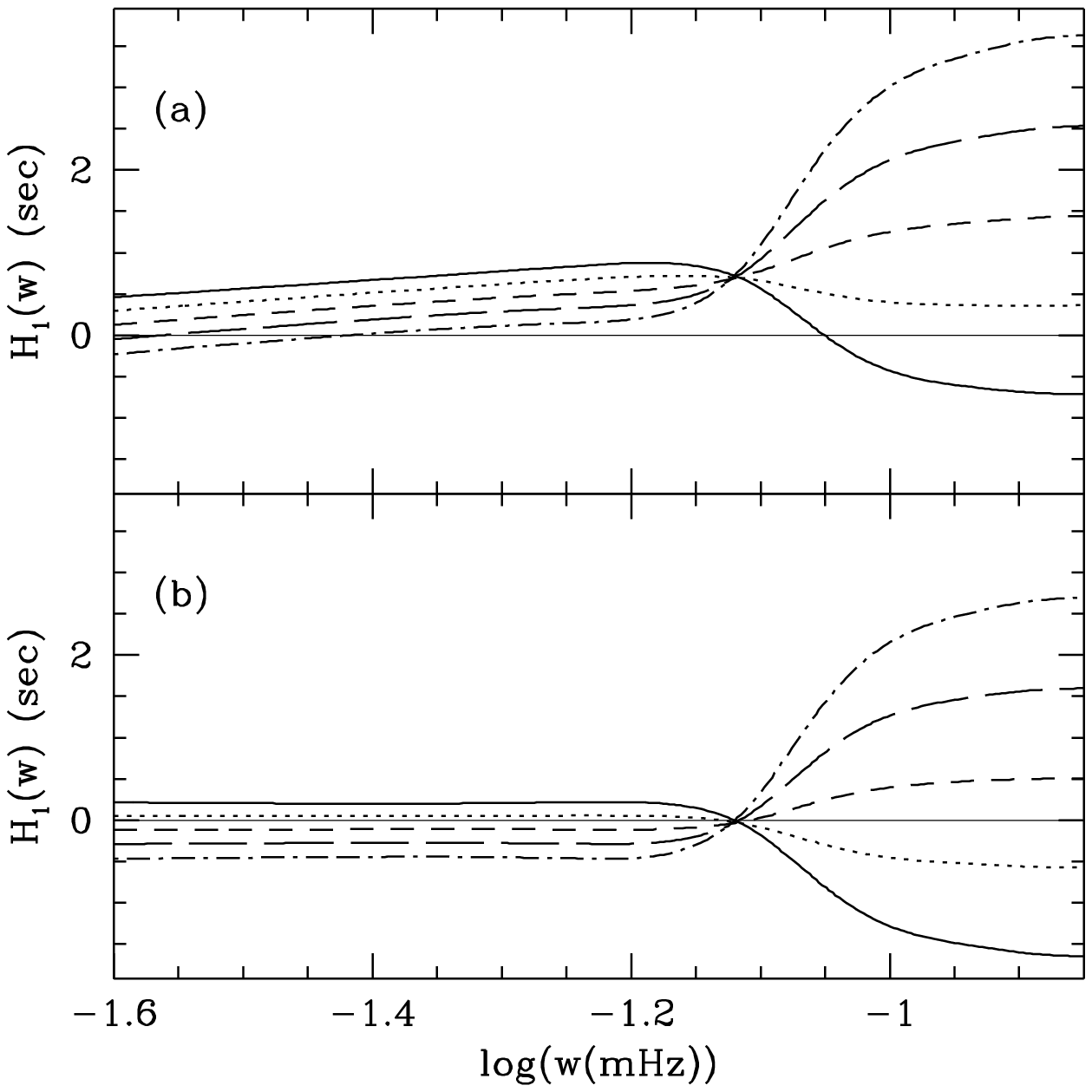}\vskip -2.25 true cm}}
\caption{\bf Fig.~5 \rm The function $H_1(w)$ between the model
INV78-0.712 and (a) the INV99 calibration models and (b) the 
INV78 calibration models. In both panels, the continuous line shows  
$H_1(w)$ between the test model and the calibration model with $r_b=0.709
R_{\rm s}$, the dotted line is with $r_b=0.711R_{\rm s}$, small-dashed
line with $r_b=0.713R_{\rm s}$, long-dashed line with $r_b=0.715R_{\rm s}$
and dot-dashed line with $r_b=0.717R_{\rm s}$.
}
\endfigure

For each type of model, we have two sets of calibration models, one with the
standard radius of 695.99 Mm and the other with the reduced radius of 
695.78 Mm. Each calibration set consists of 5 models with CZ-base
positions at $0.709$, $0.711$, $0.713$, $0.715$ and $0.717\;R_{\rm s}$.
Thus we have 6 sets of calibration models. 
In addition, we have test models for each hydrogen abundance profile and
each radius, with $r_b=0.712\;R_{\rm s}$. We also have test models
with an even lower radius of 695.57 Mm, to check for trends.
For convenience, the models are referred to by their radius (99 for 695.99 Mm,
78 for 695.78 Mm and 57 for 695.57 Mm), hydrogen and heavy element
abundance profile and
radius of the CZ base, e.g., DIF99-0.709 is the model with the DIF 
$X$ and $Z$ profiles,
with $r_b=0.709R_{\rm s}$ and standard radius ($R_{\rm s}=695.99$ Mm),
INV78-0.711 s the model with INV $X$ profile, $r_b=0.711R_{\rm s}$ and
smaller radius ($R_{\rm s}=695.78$ Mm),  etc. All models have been
constructed with  the OPAL equation of state, OPAL and Kurucz opacities
and with the CM formulation for calculating convective flux. They have
identical surface $X$ of 0.7342 and surface $(Z/X)$ of 0.0245.
The systematic errors arising due to variations in these parameters were
studied by Basu and Antia~(1997).

\subsection{Results}

\beginfigure{6}
\hbox to 0 pt{\hskip -1.5cm
\vbox to 6.5 true cm{\vskip -1.75 true cm
\epsfysize=10.50 true cm\epsfbox{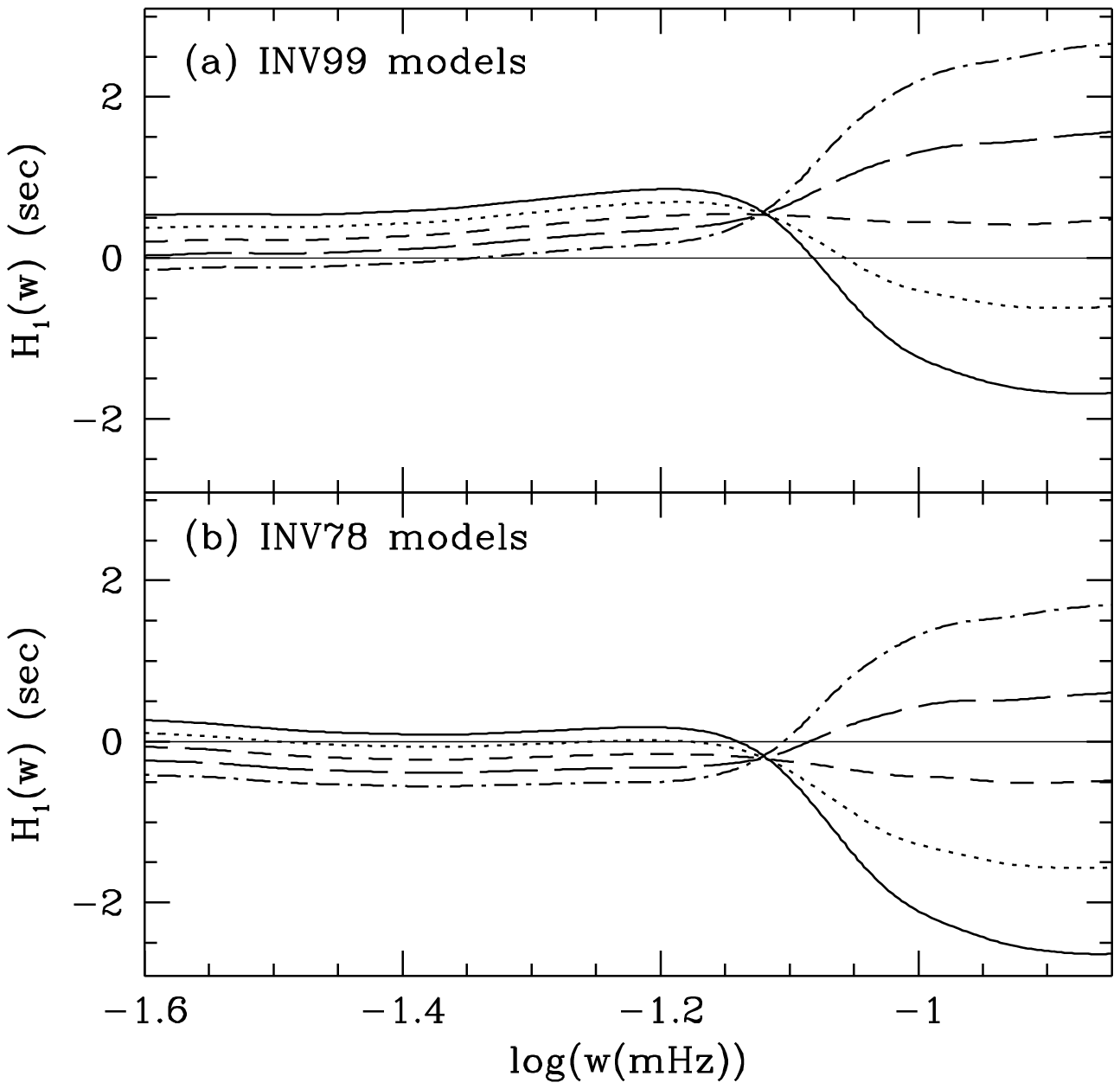}\vskip -2.25 true cm}}
\caption{\bf Fig.~6 \rm The function $H1(w)$ between the Sun and
(a) the INV99 calibration models and (b) the 
INV78 calibration models. The line types are the same as in Fig.~5.
}
\endfigure

\begintable*{2}
\caption{\bf Table 2. \rm CZ results for the Sun}
{\tablet{15.0 true cm}{#\hfil&&\hfil $#$\cr
\tabmidrule
Data set  & \multispan{3}{\hfill Deduced $r_b/R_{\rm s}$ \hfill }\cr
\tabmidrule
          & \multispan{1}{\hfil ND78 models\hfill } &
\multispan{1}{\hfil INV78 models\hfil }& 
\multispan{1}{\hfil DIF78 models \hfil}\cr
\tabmidrule
BBSO &  0.714219\pm 0.000179& 0.713435\pm 0.000175& 0.710665\pm 0.000181\cr
GONG &  0.714149\pm 0.000062& 0.713430\pm 0.000060& 0.710631\pm 0.000062\cr
MDI &   0.714443\pm 0.000080& 0.713601\pm 0.000078& 0.710868\pm 0.000081\cr
Weighted 
Average&  0.714267\pm 0.000050& 0.713494\pm 0.000049& 0.710722\pm 0.000051\cr
\tabmidrule
}
}
\endtable

The results for the test models are listed in Table~1.
We have shown the results for the INV models only. The results for the
other calibration models are similar.
 We can see that
there is indeed some difference in the measured value of radial position
of the CZ base, when reference models with a different radius are used.
The difference is of the order of the relative difference in the radius,
but is  somewhat
smaller than error in the sound speed caused by mismatch of radius.
The reason for this can be seen in Fig.~5, where we have shown the function
$H_1(w)$ for model INV78-0.712 and the INV99 and INV78 calibration models.
$H_1(w)$ is flat in the CZ for INV78 calibration models. This is consistent
with the convection-zone sound-speed difference between these models and
the test model INV78-0.712 being 0.  The predominant difference between the
results with INV99 and INV78 models is a 
linear shift. The truly linear part of the shift  is easily removed by the
function $H_s(w)$ in the fit to Eq.~\calib, hence the $r_b$ results for this
model are not very different.
Nevertheless,  the error arising due to difference in radius is larger than the
error in the solar results because of errors in the frequencies, and hence
the solar CZ depth should be determined using the models with
the correct radius.
The difference in $r_b$ due to a difference  in the radius is much 
smaller than that due to differences in the hydrogen abundance profile.
For example,
the error in determining the position of the CZ base can be 
as much as 0.0035 $R_{\rm s}$ (cf., Basu \& Antia 1997) between the
DIF and ND profiles.

The function $H_1(w)$ between the Sun and the two sets of INV calibration 
models are shown in Fig.~6. Note that $H_1(w)$ for models with 
$R_{\rm s}=695.78$ Mm is much flatter in the CZ as compared to 
models with $R_{\rm s}=695.99$ Mm. This gives added confidence that 
the solar radius is indeed close to 695.78 Mm.
Table~2 lists
the position of the solar CZ base as obtained using different data
sets and different calibration models.

Since it can be shown that models with the DIF profile have too sharp
a  gradient in the $X$-profile at the CZ base (Basu 1997a), we can 
disregard the value obtained with  those models. There is an 
enormous amount of evidence that diffusion of helium and heavy elements does
take place below the solar CZ base (see e.g. Basu 1997b), the value
obtained with the ND models is thus an upper limit to the value of 
$r_b$, as models with diffusion will give lower values.  
The value obtained with the INV models is probably the most reliable
determination. 
The change in the results is negligible if instead of the INV 
profiles, those obtained from stellar evolution 
codes that incorporate mixing due to rotation (e.g., Richard et al.~1996) 
are used.  

 The
various other systematic errors that can creep into the determination
of the position of the CZ base have been discussed in detail in 
Basu \& Antia (1997). They are invariably larger than the statistical error
due to the error in frequencies. This gives an 
error of $0.0005R_{\rm s}$, which is larger
than the error due to the uncertainty in the radius. 

The base of the convection zone is therefore  estimated to
be at  $0.7135\pm 0.0005 R_{\rm s}$. If models with the standard radius are
used, we find a position of $0.7132\pm 0.0005 R_{\rm s}$ using the
same data sets.

\beginfigure{7}
\hbox to 0 pt{\hskip -1.5cm
\vbox to 4.5 true cm{\vskip -2.9 true cm
\epsfysize=11.00 true cm\epsfbox{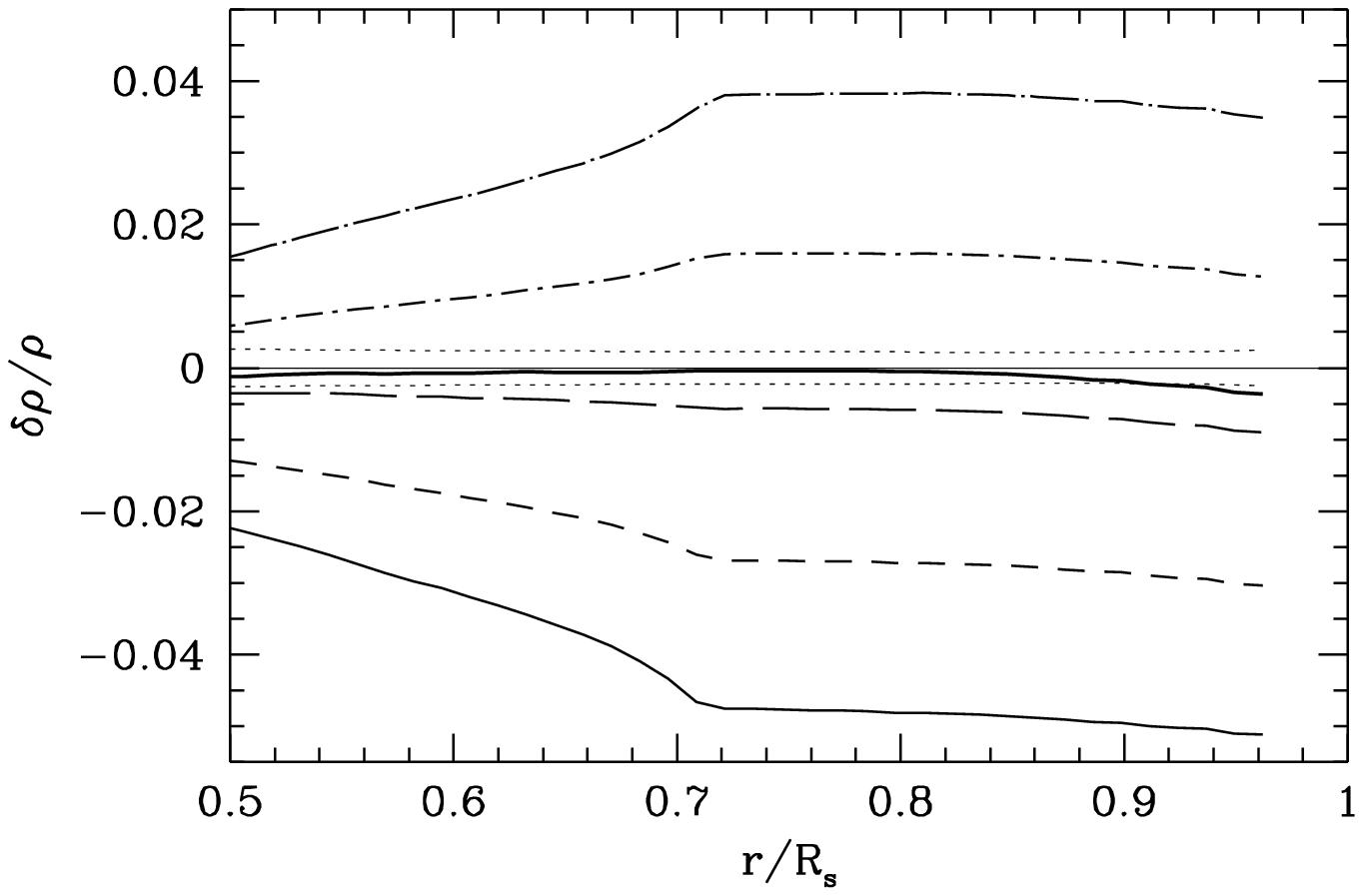}\vskip 1.00 true cm}}
\caption{\bf Fig.~7 \rm The relative density difference between the
Sun and the INV78 calibration models. The thin-continuous, small-dashed,
long-dashed, dot-small dashed and dot-long dashed lines are the
differences for the models with $r_b=$ 0.709, 0.711, 0.713, 0.715
and 0.717 $R_{\rm s}$ respectively. The thick continuous line is 
for the model with the CZ base at the position estimated using
the above calibration models ($r_b=0.7135 R_{\rm s}$). The dotted lines
show the $1\sigma$ error in the density inversion. 
}
\endfigure

\subsection{CZ depth and $Z/X$  from density differences}

The sound speed inside the convection zone is not particularly sensitive
to the composition or the position of the convection-zone base. 
The difference in the sound speed between various models occurs
below the CZ base.
Thus it is difficult to disentangle the change
in sound-speed below the CZ base because of change in temperature
gradient from that due to changes in the mean-molecular weight
due to gravitational settling of helium and heavy elements.
This is the reason that the models with different composition
profiles give seemingly different estimates of the position
of the CZ base. The density in the CZ is, however,
affected by the position of the CZ base, which arises because of difference
in mixing length required to achieve different depths of CZ.
This can be seen from 
Fig.~7, where we show the density difference between Sun and 
 the INV calibration models. We have also shown the density difference
for the model with the CZ base at the deduced position of 
$0.7135R_{\rm s}$. Note that the density difference for this model
is consistent with 0, which gives us added confidence in the 
results. 
It may be noted that the position of the base of the convection zone is
determined by the opacity and in principle, it is possible to estimate the
opacity at the CZ base by using the inversion results coupled with an
estimate of the position of the CZ base. 
Analyses using density differences unfortunately suffer
from the drawback that density in the CZ is also dependent on the
low-temperature opacities used and to some extent on the composition
within the CZ.

As can be seen from Fig.~7, the density difference between the
INV78 model with $r_b=0.7135$ and surface $Z/X=0.0245$ and the Sun
is almost zero. This would mean that assuming that $r_b$ and $Z/X$
are correct, the opacities used in the model, i.e., OPAL supplemented
by Kurucz,  are consistent with helioseismic data.
The uncertainty in the estimated 
depth of the CZ base  i.e., 0.0005$R_{\rm s}$ corresponds to a 
change of  opacity at the CZ base by 0.6\%, which is much smaller than
the probable uncertainties in the opacity. The
largest change in opacity occurs when $Z/X$ is changed. If $Z/X$ is
changed by the quoted uncertainty of 10\%, the opacities also show
almost 10\% change. Thus we can conclude that with the current
values of $Z/X$ the OPAL opacity tables are consistent with the
estimated opacity at the base of the solar CZ.

The good agreement between the estimated opacity and that calculated
from the OPAL tables using the currently accepted value of $Z/X$ 
suggests that the uncertainty in $Z/X$ should be much smaller than the
quoted error of 10\%.
Assuming  the correctness of the OPAL opacities  we can use 
the CZ-density differences
between the Sun and the models to put limits on the amount of heavy
elements in the solar envelope. The solar models constructed with
$r_b=0.7135R_{\rm s}$, $Y=0.2478$ and $Z/X=0.0245$ shows a near
perfect fit to the CZ density. If we keep $Y$ the same, then we
find that a change in $Z/X$ of $\pm 0.0001$ changes the density
difference by $\pm 1\sigma$. If we keep $X$ fixed and change
$Z/X$ then a change of $\pm 0.0002$ changes density by $\pm 1\sigma$.
In addition there are uncertainties caused by errors in the depth of the
CZ, the low temperature opacities, the hydrogen abundance in the CZ,
and uncertainties in the solar-radius.
%
Taking into account all these uncertainties
we estimate that the  heavy element abundance in the solar 
envelope is $Z/X=0.0245\pm 0.0006$, i.e., the uncertainty is
much less than the normally assumed uncertainty of 10\% in spectroscopically
estimated value.

\subsection{Overshoot below the solar CZ base}

\begintable{3}
\caption{\bf Table 3. \rm Amplitude of signal form CZ base}
{\tablet{8.0 true cm}{#\hfil&&\hfil $#$\cr
\tabmidrule
Model  & \multispan{2}{\hfill Amplitude ($\mu$Hz)\hfill }\cr
& \hbox{INV99}  &  \hbox{INV78} \cr
\tabmidrule
Ov.=0 $H_p$ & 0.8020 & 0.8073\cr
Ov.=0.05 $H_p$ &0.9014&  0.9071\cr
Ov.=0.10 $H_p$ &1.0428 & 1.0494 \cr
\tabmidrule
 \multispan{3}{\hfill Observations \hfill }\cr
\tabmidrule
BBSO & \multispan{2}{\hfill $0.879\pm 0.106$\hfill } \cr
GONG4-14 & \multispan{2}{\hfill $0.782\pm 0.032$\hfill }\cr
MDI &  \multispan{2}{\hfill $0.771\pm 0.034$\hfill }\cr 
Weighted  Average& \multispan{2}{\hfill $0.790\pm 0.025$\hfill }\cr
\tabmidrule
}
}
\endtable

Solar oscillation frequencies can be used to determine the extent
of overshoot below the solar convection zone if it is assumed that the
overshoot region is adiabatically stratified and that there is an
abrupt transition to radiative stratification below the overshoot
zone.
This change in stratification introduces a discontinuity in the first
derivative of the sound-speed and leaves a characteristic oscillatory
signal in the frequencies. The amplitude of the signal can be 
calibrated to determine the extent of overshoot
(cf., Monteiro, \jcd\ \& Thompson 1994; Basu \& Antia 1994b; Basu 1997a).
Basu (1997a) had found an upper limit of $0.05H_p$ on the overshoot
region. That was done using models with the standard radius of 
695.99 Mm. We have tried to see whether models with the lower
radius show a change in the amplitude of the signal and whether
the use of the new data changes the limits. The procedure used
for fitting is described in Basu (1997a).

We constructed a few models to test  the change of the amplitude
with radius and extent of overshoot. All models are INV models.
The CZ depth of the models is appropriate for the radius of the
model.
The results for the different models and data sets are shown in Table~3.
The addition of the two data sets does not change the observed
results significantly. 
The change in the amplitude of the signal with change in 
radius is smaller than the statistical error in the amplitude
due to uncertainties in observed frequencies. 
Hence the conclusions about the extent
of overshoot below the solar convection zone are not affected by
possible error in solar radius.

\section{Helium abundance}

Spectroscopic measurements of the abundance of helium in the Sun being
very uncertain, helioseismology plays a major role in determining
the helium abundance in the solar envelope. The abundance is
obtained from the variation of the adiabatic index of the solar
material in the helium ionisation zone, which in turn
affects the sound speed.

\subsection{The procedure}

The technique used here is the same as that in
Antia \& Basu (1994b) and Basu \& Antia (1995), and is very similar to that
used to determine the depth of the CZ zone.
The starting point of the procedure is again Eq.~\dif. 
However, since
the data do not extend to high degrees and the data sets have hardly any
modes
with turning points in the helium ionisation zone ($r >  0.98 R_{\rm s}$),
it is difficult to use $H_1(w)$ to determine the helium abundance.
We have to use the
function $H_2(\omega)$ instead. It has been shown that
(e.g., Basu \& Antia 1995; P\'erez Hernandez \& \jcd\ 1994), the helium 
ionisation zone being sufficiently close to the solar surface leaves 
its imprint on the function $H_2(\omega)$ also, in the form of a hump 
around 2.5 mHz, which can be calibrated to determine the helium abundance.
The effect of difference in other
physical properties, like the equation of state, manifests itself as a
smooth curve on which the hump due to difference in helium abundance
is superimposed.

Thus, analogous to Eq.~\calib, we write the function $H_2(\omega)$ between the
Sun and a calibration model of known helium abundance as:
$$
H_2(\omega)=\beta\Phi(\omega)+H_s(\omega),\eqno\eqname\calibo$$
where $\Phi(\omega)$ is the difference in $H_2(\omega)$ between two models
which differ only in the abundance of helium. 

\begintable{4}
\caption{\bf Table 4. \rm Helium abundance results for the test models}
{\tablet{8.4 true cm}{#\hfil&&\hfil $#$\cr
\tabmidrule
Test model & \hbox{Exact} Y &\multispan{2}{\hfil Deduced $Y$ \hfil }\cr
           &&    \hbox{$R_{\rm s}=695.99$ Mm} & \hbox{$R_{\rm s}=695.78$ Mm} \cr
\tabmidrule
          && \multispan{2}{\hfil Calib. models with OPAL EOS\hfil }\cr
\tabmidrule
OPAL99-0.73& 0.252 & 0.25184 & 0.25156 \cr
OPAL78-0.73 & 0.252 & 0.25212 & 0.25184 \cr
OPAL57-0.73& 0.252 & 0.25240 & 0.25212 \cr
MHD78-0.73 & 0.250& 0.24938 & 0.24904   \cr
\tabmidrule
          & &\multispan{2}{\hfil Calib. models with MHD EOS\hfil }\cr
\tabmidrule
MHD99-0.73 & 0.250 & 0.24987& 0.24957\cr
MHD78-0.73  & 0.250 & 0.25016 & 0.24987\cr
MHD57-0.73 & 0.250 & 0.25046& 0.25016\cr
OPAL78-0.73 & 0.252 & 0.25373 & 0.25355 \cr
\tabmidrule
}
}
\endtable

\begintable{5}
\caption{\bf Table 5. \rm Helium abundance  results for the Sun}
{\tablet{8.4 true cm}{#\hfil&&\hfil $#$\cr
\tabmidrule
Data set  & \multispan{2}{\hfill Deduced $Y$\hfill }\cr
           & \hbox{$R_{\rm s}=695.99$ Mm} & \hbox{$R_{\rm s}=695.78$ Mm} \cr
\tabmidrule
          & \multispan{2}{\hfil OPAL models\hfil }\cr
\tabmidrule
BBSO & 0.2476\pm 0.0002&  0.2474\pm 0.0002\cr
GONG & 0.2471\pm 0.0001& 0.2470\pm 0.0001\cr
MDI & 0.2488\pm 0.0001 & 0.2488\pm 0.0001\cr
\tabmidrule
          & \multispan{2}{\hfil MHD models\hfil }\cr
\tabmidrule
BBSO& 0.2514\pm 0.0002 & 0.2510\pm 0.0002\cr
GONG & 0.2499\pm 0.0001& 0.2496\pm 0.0001\cr
MDI &  0.2524\pm 0.0001& 0.2522\pm 0.0001\cr
\tabmidrule
}
}
\endtable

For this work too it is preferable to use envelope models since the calibration
models are required to have specified values of the helium abundance.
Since by far the largest error in helium abundance determination
occurs due  to a mismatch of the equation of state (cf., Antia \& Basu
1994b), we use models with both OPAL and the 
MHD (Hummer \& Mihalas 1988; Mihalas, D\"appen \& Hummer 1988;
D\"appen \etal~1988) equations of state. For each equation
of state, we have constructed calibration sets with radii 695.99 Mm 
and 695.78 Mm. Each calibration set consists of five models with 
hydrogen abundance $X$ of 0.68, 0.70, 0.72, 0.74 and 0.76. The
value of the metal abundance $Z$ was kept fixed at $0.018$ for the OPAL
models. For the MHD models, the available tables force $Z=0.02$.
All models have the CZ base at $r_b=0.713\;R_{\rm s}$.
In addition, we have test models with $X=0.73$ for both equations
of state and both values of the radius. We also have constructed test
models with $R_s=695.57$ Mm. The models are labelled by their
EOS (OPAL or MHD),
radius (99 for 695.99 Mm, 78 for 695.78 Mm and 57 for 695.57 Mm),
 and surface hydrogen abundance. 
Since the helium
ionisation zone is very far from the CZ base,
it is not expected to be affected by the abundance
profiles below the CZ base. Hence, the models constructed for calibrating
the helium abundance do not
assume any diffusion of helium or heavy elements below the CZ base.

\subsection{Results}

The value of $Y$ obtained for different models are listed in Table~4.
Note that unlike  the case of the CZ depth, where the CZ depth of the
models with same $R_{\rm s}$ could be determined accurately,  there is
an  error of about 0.0002 in the determination of the helium abundance 
even when 
the radius is the same. 
The main reason is that the signature due to differences in  helium abundance
is not as clear-cut as that due to differences in the position of the
CZ base. Another reason for this is that the mode-sets
used do not have modes with turning points in the helium ionisation
zone. Nonetheless, we see that there is an additional error if the 
radius of the calibration and test models differs. The magnitude of the
difference is of the order of  the fitting errors in the determination of 
helium abundance. However, as we shall show below, there are larger
errors caused by a number of other factors, such as opacity used etc.
Hence the solar results are unlikely to be
affected significantly by a change in radius of the calibration models.

Since the contribution to $H_2(\omega)$ at any given frequency comes from
modes with different turning points, the deduced helium abundance 
may be affected by model differences not necessarily localised 
at the helium ionisation zone. To test this we have constructed a few
additional models to test the error in the deduced helium
abundance caused because of  differences  in composition profile
below the CZ base, depth of CZ and opacities used. As expected, the
composition profile below the CZ base does not affect the determination
of He abundance.
However, the depth of the CZ does. An error of $0.0005R_{\rm s}$ in the
CZ depth causes an error of 0.0003 in $Y$. A much larger error
of 0.0006 is produced if the low temperature Kurucz (1991) opacities
are replaced by those of Alexander (1975).
We estimate that the 
 uncertainty in the deduced helium abundance due to all these
factors is about 0.001.

The results for the Sun are listed in Table~5.
 The error-estimate corresponds
merely to the error in the result due to those in the frequencies, and as
can be seen in light of the discussion above, are much smaller than the
systematic errors involved.
Unlike in the case of CZ depth determination, the different data sets
give results that  are not consistent with each other within the errors.
We believe this is a result of the different degree  ranges of the three
data sets. For a definitive determination of the helium abundance, one
needs high-degree p-modes which have lower turning points in the 
helium ionisation zones. None of the data-sets used has such high degree
modes. The highest-degree p-mode in the MDI set has ${\ell}=194$, 
in the GONG set it has $\ell=150$ and the BBSO set has been deliberately
 restricted to modes of ${\ell} \le 140$
to avoid using modes with large systematic errors. Since none of the
mode-sets has modes which sample the helium ionisation zone very well,
it is not very surprising  that the results with the three data sets are
different

As expected (cf., Basu \& Antia 1995; Kosovichev 1996; Baturin \& Ayukov 1997),
MHD and OPAL models give different results for the Sun.   The result obtained
for the OPAL models, an average of the three sets being  $Y=0.2478$, is not 
very different from the earlier
result of $Y=0.249$ obtained by Basu \& Antia (1995), and are the same as
those of Kosovichev (1996). 
The current results are, however, expected
to be more reliable since the radius of the models is consistent with
that of the Sun and the data-sets used are better. For models with the
standard radius of $695.99$ Mm, we find $Y=0.2479$, with an uncertainty
of $0.001$ in the results due to fitting errors and systematic 
error mentioned above.

The result for the MHD models on the other hand is somewhat
higher than that obtained
by Basu \& Antia (1995) but are consistent with those of Baturin \& Ayukov
(1997).  The results obtained by different workers using MHD models have 
a wide dispersion. This may be due to the fact that the sound-speed difference
between all MHD models and the Sun show a distinct deviation just
below the helium ionisation zone (cf., Dziembowski, Pamyatnykh \&
Sienkiewicz 1992;
Basu \& Antia 1995), which can contaminate the signal from the helium
abundance to varying degrees depending on the procedure followed to
estimate the helium abundance in the solar envelope.

\section{ The sound-speed difference in the CZ}

\beginfigure{8}
\hbox to 0 pt{\hskip -1.5cm
\vbox to 6.5 true cm{\vskip -1.75 true cm
\epsfysize=10.50 true cm\epsfbox{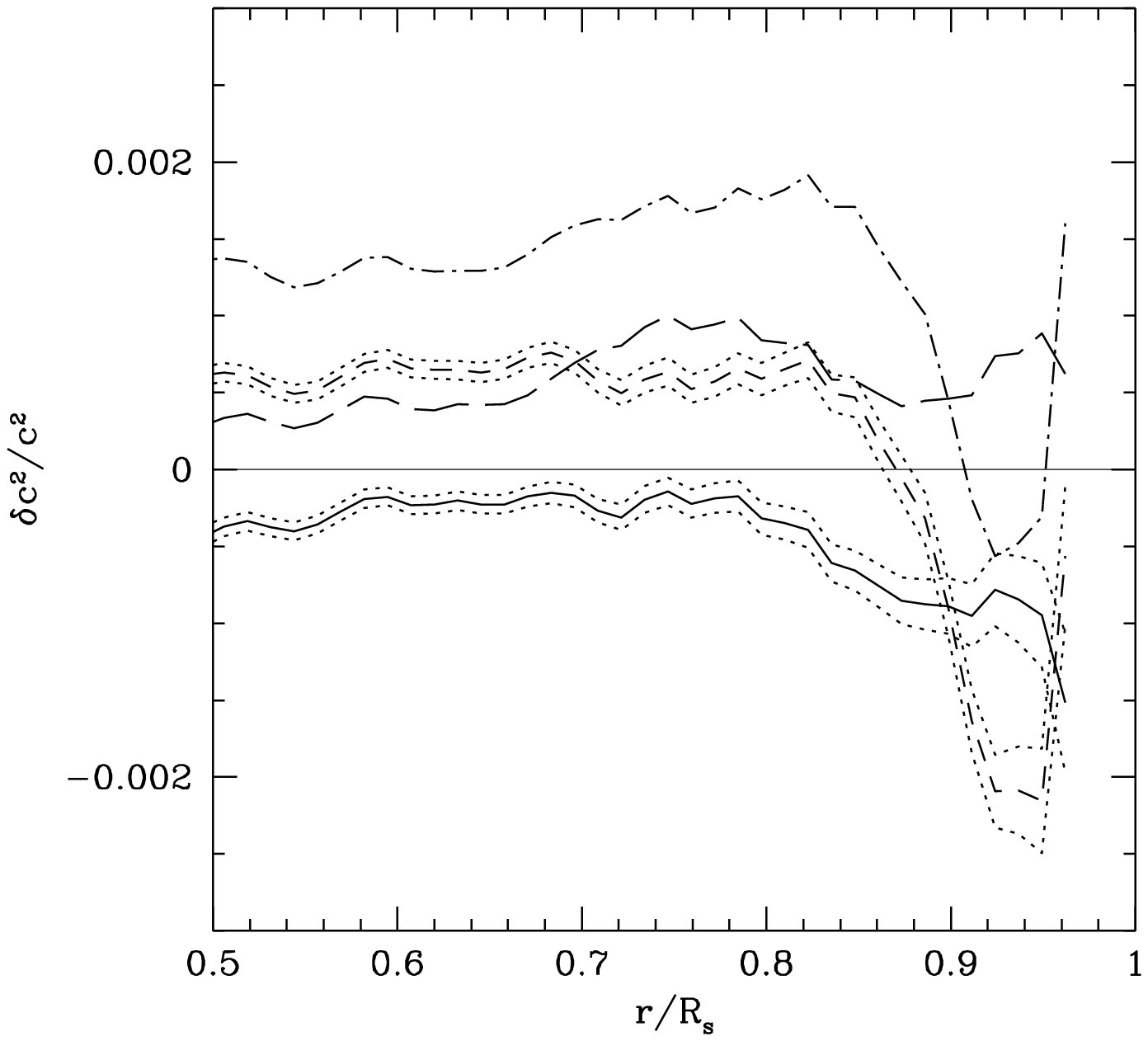}\vskip -2.25 true cm}}
\caption{\bf Fig.~8 \rm The relative sound-speed difference between 
solar envelope models and the Sun. The continuous line  and 
the small-dashed line 
are for models  with $R_{\rm s}=695.78$ Mm, constructed with OPAL 
and MHD equations of state respectively. The dotted lines are the $1\sigma$
error envelope. The long-dashed and the dot-dashed lines are for
models with $R_{\rm s}=695.99$ Mm,  constructed with OPAL 
and MHD equations of state respectively. For the sake of clarity, the
error envelopes have not been shown for these models.
}
\endfigure

Fig.~8 shows the sound-speed difference between the Sun and four 
solar models in the convection zone.  There are two OPAL models and two 
with the MHD EOS. 
For each EOS, one model has the standard radius of 695.99 Mm and one the
reduced radius of 695.78 Mm. Each model has a CZ-base position and helium
abundance appropriate to it's radius and EOS (cf., Sections 3 and 4).
Note that the  sound-speed of the Sun is higher than that of the 
two models with the standard radius. For the reduced-radius models,
the solar sound-speed is higher than the sound-speed of the MHD model,
but is marginally lower than that of the OPAL model. The overall
difference is however much lower than the models with the standard 
radius.

\beginfigure{9}
\hbox to 0 pt{\hskip -1.5cm
\vbox to 4.5 true cm{\vskip -3.00 true cm
\epsfysize=11.50 true cm\epsfbox{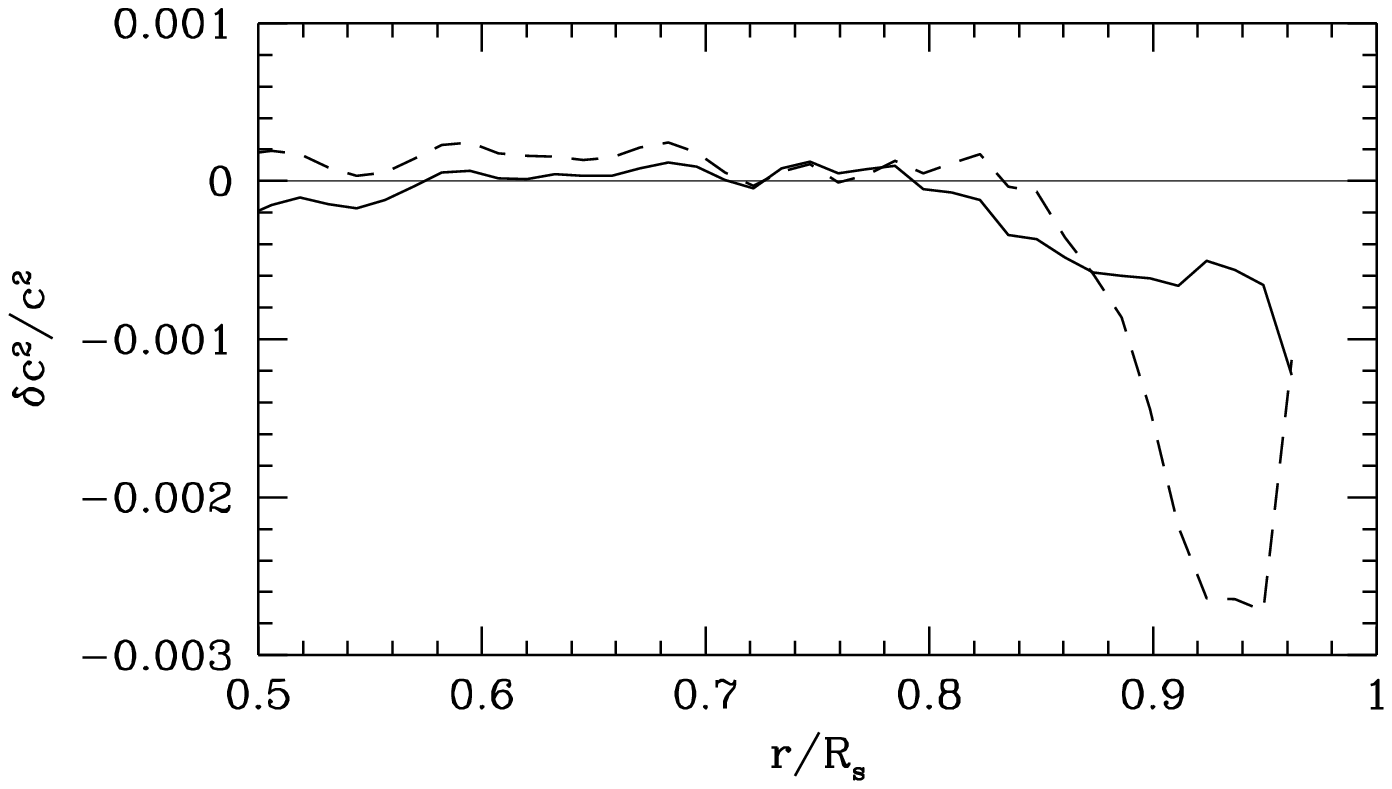}\vskip -2.25 true cm}}
\caption{\bf Fig.~9\rm The relative sound-speed difference between 
solar envelope models and the Sun. Both models have a radius of 695.78 Mm.
The continuous line is for a model  constructed with a modified OPAL
EOS, and the dashed line is for a model constructed with modified
MHD EOS.
}
\endfigure

There are a number of  possibilities as to why the sound-speed in the lower
part of the CZ
in the models do not match that of the Sun, these are, for instance, that the
change in radius has not been estimated correctly or that  the equation of 
state is not correct. 

Looking at Fig.~8, it appears that the change in radius has over-corrected
the sound-speed in the case of the OPAL model, while the correction
is not enough for the MHD models.  Thus, 
for the OPAL models the reduction of radius to 695.78 Mm is an
overestimate of the reduction, while for the MHD models it is an 
underestimate. The radii at which the  models show very little 
difference with
respect to the solar sound speed  in the lower CZ are 695.82 Mm and 695.67 Mm
respectively for the OPAL and MHD EOS.
This additional change in radius, however, causes the f-mode frequencies of the
model to deviate from the observed frequencies. This means that either
the surface layers of the solar models have not been modelled properly,
or other processes, like flows etc., may need to be postulated to explain the
f-mode frequency difference.

The structure of lower part of the solar convection zone is determined
mainly by the 
equation of state, hence an error in the equation of state will
 give rise to a sound-speed difference between a solar model and the Sun.
Thus if we assume that the radius and helium abundance in CZ have been
estimated correctly, the remaining discrepancy in the sound speed may be
due to possible error in the equation of state.
It is clear from Fig.~9 that the OPAL EOS
is much closer to that of the solar material than the MHD EOS. However,
there is still some remaining difference. We find that for the reduced
radius models, a large part of the difference
can be eliminated if the partial derivative
$(\partial p/\partial T)_\rho$ for the OPAL
equation of state is lowered by a factor $(1-0.00027)$, and the
other thermodynamic quantities like $C_p$ and $\Gamma_1$ are
changed consistently.  This corresponds
to a relative difference of about $-0.00022$ in   $\Gamma_1$
between the two models in the lower part of the CZ. For the
MHD EOS, the partial-derivative should be larger  by a factor 1.00053, this
results in a relative difference of 0.00043 in $\Gamma_1$. This small
difference largely eliminates the sound-speed difference in the CZ, 
as can be seen from Fig.~9. It is not obvious that the discrepancy is
indeed due to an error in equation of state, but these differences
probably give an estimate of uncertainties in current equations of state.
This change in EOS does not change the f-mode
frequencies significantly.

\section{Conclusions}

We have shown that a mismatch in the radius of the Sun and that of reference
solar models can introduce considerable error in the inverted sound-speed
and density profiles of the Sun.  The relative difference in the 
sound speed obtained by inversion due to the change in radius is
of the same order as the relative difference between the sound speed
of the Sun and  the solar models used.
Therefore, inversion of solar 
oscillation frequencies to determine solar structure should be performed
using models with radius as close as possible to the correct value.
Thus it is necessary to obtain an accurate estimate of the solar radius.

We find that regardless of the radius of the solar models used, solar
models constructed using the CM formulation of convection  give a better
fit to the near-surface solar sound speed  than models constructed
with the mixing length formalism. Models constructed with the
mixing length formalism show a sharp dip in the relative sound-speed
difference with respect to the Sun in the outer  regions, and this dip 
is not removed by the reducing the radius of the solar models.

We have used new data-sets to check how 
the change in radius results in a small change in the measured depth
of the solar convection zone and in the helium abundance. We 
obtain a radial distance of $(0.7135\pm 0.0005)R_{\rm s}$ for the base 
of the CZ. 
We also find that the OPAL tables are consistent
with the solar opacities at the CZ base for a surface $Z/X$ of
0.0245. Assuming that the OPAL opacities are correct, we deduce that the
the surface $Z/X=0.0245\pm0.0006$ with an error estimate
which is much less than the usually quoted value of 10\%.
The limit on the  extent of overshoot does not change because of the
small change in radius, and remains at $0.05H_p$ (2800 Km).

Using the new data sets and models constructed
with the OPAL equation of state, we find that the  helium abundance 
in the solar envelope is  $0.2478 \pm 0.001$. 
Once again, the estimate of helium abundance is not significantly affected
by the error in radius as the effect is smaller than the systematic
errors in this measurement.

If the correct $R_{\rm s}$ is indeed 695.78 Mm and abundances in CZ
have been estimated correctly,
then the remaining discrepancies between the CZ sound-speed
profile of the models and the Sun can be explained as errors in the
equation of state used.
It can be shown that
by changing $(\partial p/\partial T)_\rho$ for each EOS suitably,
one can reduce the difference. For OPAL models this has to be
reduced by a factor of  0.99973, while for MHD models this
has to be increased by a factor 1.00053. Changes in 
$(\partial p/\partial T)_\rho$ automatically changes the
adiabatic index $\Gamma_1$, and corresponds  to a decrease in
$\Gamma_1$ by a factor of 0.99978 for OPAL models and an
increase by a factor of 1.00043 for MHD models. This probably gives an
estimate of uncertainties in the current equations of state.

\section*{Acknowledgments}

This work utilizes data obtained by the Global Oscillation
Network Group (GONG) project, managed by the National Solar Observatory, a
Division of the National Optical Astronomy Observatories, which is 
operated by AURA, Inc. under a cooperative agreement with the 
National Science Foundation. The data were acquired by instruments 
operated by the Big Bear Solar Observatory, High Altitude Observatory,
Learmonth Solar Observatory, Udaipur Solar Observatory, Instituto de 
Astroflsico de Canarias, and Cerro Tololo Inter-American Observatory.
SOHO is a
project of international cooperation between ESA and NASA
The author was supported by an AMIAS fellowship and the Institute for
Advanced Study SNS Membership fund.


\section*{References}


\beginrefs 

\bibitem Alexander D. R., 1975, ApJS, 29, 363

\bibitem Antia  H. M., 1997, A\&A, in press (astro-ph/9707226)

\bibitem Antia H. M., Basu S., 1994a, A\&AS, 107, 421

\bibitem Antia H. M., Basu S., 1994b, ApJ, 426, 801

\bibitem Antia H. M., Basu S., 1997,  in  eds. Pijpers F. P., \jcd\ J.,
Rosenthal C.,  Solar Convection and their Relationship,
(Dordrecht: Kluwer), p51

\bibitem Antia  H. M., Chitre S. M., 1997a   (astro-ph/9710159)

\bibitem Antia  H. M., Chitre S. M., 1997b,  in Proc: IAU Symp. 185, in
press

\bibitem Allen C. W. 1976, Astrophysical Quantities (3d ed.; London: Athlone)

\bibitem Bahcall, J. N.,  \& Pinsonneault, M. H., 1992, Rev. Mod. Phys.~64, 885

\bibitem Balmforth N. J., 1992, MNRAS 255, 632

\bibitem Basu S., 1997a, MNRAS, 288,  572

\bibitem Basu, S., 1997b, in eds.,  Provost, J., Schmider, F. -X.,
IAU Symp. 181: Sounding Solar and Stellar Interiors,
Kluwer, Dordrecht, in press

\bibitem Basu S., Antia H. M., 1994a, JAA, 15, 143

\bibitem Basu S., Antia H. M., 1994b, MNRAS, 269, 1137

\bibitem Basu S., Antia H. M., 1995,  MNRAS,  276, 1402


\bibitem Basu S., Antia H. M., 1997,  MNRAS,  287, 189

\bibitem Basu S., Christensen-Dalsgaard J., Schou J., Thompson M. J.,
Tomczyk S. 1996, ApJ, 460, 1064

\bibitem
Baturin V. A., Ayukov S. V, 1997, in  eds. Pijpers F. P., \jcd\ J., 
Rosenthal C.,  Solar Convection and their Relationship,
(Dordrecht: Kluwer), p55

\bibitem Canuto, V. M., Mazzitelli, I, 1991, ApJ 370, 295

\bibitem Christensen-Dalsgaard, J., Gough, D. O. and Thompson, M. J. 1989,
MNRAS 238, 481

\bibitem \jcd\ J., Gough, D. O. \& Thompson, M. J. 1991, ApJ 378, 413

\bibitem \jcd\ J., D\"appen W., Ajukov S. V., et al., 1996,  Science,
272, 1286

\bibitem Cox A. N., Kidman R. B., 1986, in Theoretical Problems in 
Stellar Stability and Oscillations (Institu d'Astrophysiquw, Li\'ege)
p259

\bibitem D\"appen W., Mihalas D., Hummer D. G., Mihalas B. W., 1988,
 ApJ,  332, 261

\bibitem D\"appen W., Gough D. O., Kosovichev A. G., Thompson M. J.,
 1991, in eds.,  Gough D. O.,  Toomre J., 
 Lecture Notes in Physics, 388, Springer, Heidelberg, p.111


\bibitem Dziembowski W. A., Pamyatnykh A. A., Sienkiewicz R., 1990,
MNRAS, 244, 542

\bibitem
Dziembowski W. A., Pamyatnykh A. A., Sienkiewicz R., 1992,
 Acta Astron.,  42, 5

\bibitem Dziembowski W. A., Goode P. R., Pamyatnykh A. A., Sienkiewich R.,
1994,  ApJ,  432, 417

\bibitem
Elsworth Y., Howe R., Isaak G. R., et al., 1994,  ApJ,  434, 801


\bibitem Gough D. O., Kosovichev A. G., Toomre J., et al., 1996,  Science,
 272, 1296

\bibitem Grevesse N.,  Noels A. 1993,
eds.  Prantzos N., Vangioni-Flam E.,  \&  Cass\'e M.,
in  Origin and evolution of the Elements,
(Cambridge: Cambridge Univ. Press), p15

\bibitem Guzik, J. A.,  Cox, A. N., 1993, ApJ 411, 394

\bibitem Hill F., Stark P. B., Stebbins R. T. et al.,~1996,
Science, 272, 1292

\bibitem Hummer D. G., Mihalas D., 1988,  ApJ,  331, 794

\bibitem Iglesias, C. A. \& Rogers, F. J. 1996, ApJ 464, 943

\bibitem Kosovichev A. G., 1996, Bull. Astron. Soc. India,  24, 355

\bibitem  Kosovichev, A. G., Fedorova, A. V., 1991, Sov. Astron., 35, 507

\bibitem Kosovichev A. G., Schou J., Scherrer P., et al., 1997,
Solar Phys., 170, 43

\bibitem Kurucz R. L., 1991, in eds., Crivellari L., Hubeny I., Hummer D.G.,
 NATO ASI Series, Stellar Astmospheres: Beyong Classical Models.
Kluwer, Dordrecht, p.441

\bibitem Libbrecht K. G., Woodard M. F., Kaufman J. M., 1990,  ApJS,  74, 11
29

\bibitem Mihalas D., D\"appen W., Hummer D. G., 1988,  ApJ,  331, 815

\bibitem Monteiro M. J. P. F. G., \jcd\ J., Thompson M. J., 1994,
A\&A, 283, 247

\bibitem P\'erez Hern\'andez F., \jcd\ J., 1994,  MNRAS,  267, 111

\bibitem Pijpers  F. P., Thompson M. J.,  1992,  A\&A,  262, L33

\bibitem Proffitt C. R. 1994, ApJ 425, 849

\bibitem Richard O., Vauclair S., Charbonnel C., Dziembowski W. A.,  1996,
 A\&A,
312, 1000

\bibitem Rhodes E. J., Kosovichev A. G., Schou J., Scherrer P. H.,
Reiter J., 1997, Solar Phys., 175, No:2 in press

\bibitem Rogers, F. J., Swenson, F. J., Iglesias, C. A., 1996, ApJ 456, 902

\bibitem Schou J., Kosovichev A. G., Goode P. R., Dziembowski W. A,
1997, ApJ, 489, L197

\bibitem Thompson M. J., Toomre J., Anderson E. R., et al.,
1996, Science, 272, 1300

\bibitem Vernazza J. E., Avrett E. H., Loeser R., 1981,
ApJS, 45, 635


\endrefs
\bye